\documentclass[aps, prl, twocolumn, superscriptaddress]{revtex4-1}
\usepackage[latin1]{inputenc} 
\usepackage[T1]{fontenc}

\usepackage{graphicx}
\usepackage{amsmath,amssymb,color,bm}
\usepackage{mathrsfs}
\usepackage[dvipsnames]{xcolor}

\renewcommand{\S}{\mathbf{S}}
\renewcommand{\d}{\mathrm{d}}
\newcommand{\h}{\mathbf{h}}
\newcommand{\del}{\boldsymbol{\nabla}}
\renewcommand{\P}{\mathbb{P}}

\newcommand{\CCQ}{Center for Computational Quantum Physics, Flatiron Institute, 162 5th Avenue, New York, NY 10010, USA}
\newcommand{\rev}[1]{{\color{black} {#1}}}

\begin{document}
\title{Exact numerical solution of the \rev{fully-connected} classical and quantum \\ Heisenberg spin glass}
\author{Nikita Kavokine}
\affiliation{\CCQ}
\affiliation{Department of Molecular Spectroscopy, Max Planck Institute for Polymer Research, Ackermannweg 10, 55128 Mainz, Germany}
\author{Markus M\"uller}
\affiliation{Paul Scherrer Institut, CH-5232 Villigen PSI, Switzerland}
\author{Antoine Georges} 
\affiliation{
Coll\`ege de France, Universit\'e PSL, 11 place Marcelin Berthelot, 75005 Paris, France}
\affiliation{\CCQ}
\affiliation{Centre de Physique Th\'eorique, Ecole Polytechnique, CNRS, Institut Polytechnique de Paris, 91128 Palaiseau Cedex, France}
\affiliation{DQMP, Universit\'{e} de Gen\`{e}ve, 24 quai Ernest Ansermet, CH-1211 Gen\`{e}ve, Suisse}
\author{Olivier Parcollet}
\affiliation{\CCQ}
\affiliation{Universit\'e Paris-Saclay, CNRS, CEA, Institut de Physique Th\'eorique, 91191, Gif-sur-Yvette, France}

\date{\today}

\begin{abstract} 

We present the mean field solution of the quantum and classical Heisenberg spin glasses, 
using the combination of a high precision numerical solution of the Parisi full replica symmetry breaking 
equations and a continuous time Quantum Monte Carlo.
We characterize the spin glass order and its low-energy excitations down to zero temperature.
The Heisenberg spin glass has a rougher energy landscape than its Ising analogue, 
and exhibits a very slow temperature evolution of its dynamical properties.
We extend our analysis to the doped, metallic Heisenberg spin glass,
which displays unexpectedly slow spin dynamics, reflecting the proximity to the melting quantum critical point and its associated Sachdev-Ye-Kitaev Planckian dynamics.

\end{abstract}
\maketitle 

%%%%%%%%%%%%%%%%%%%%%%%%%%%%%%% Introduction %%%%%%%%%%%%%%%%%%%%%%%%%%%%%%%
While the physics of classical and quantum Ising spin glasses has been rather thoroughly understood, glasses of Heisenberg (vector) spins have remained a difficult and largely unsolved problem, including especially its quantum version, which governs the local moments in randomly doped, strongly correlated materials.

The approach of Sachdev and Ye~\cite{sachdev_gapless_1993} (SY), which takes a
double limit of a fully connected exchange model with $\mathrm{SU}(2)$ spins
promoted to $\mathrm{SU}(M\gg 1)$, 
has attracted a lot of attention, as it
exhibits very interesting features in the 
$M=\infty$ limit. In a fermionic representation of the $\mathrm{SU}(M\to
\infty)$ generators, the spins do not freeze, but remain in a spin liquid
state~\cite{sachdev_gapless_1993} similarly to the  
related SYK model of randomly coupled Majorana fermions~\cite{chowdhury_sachdev-ye-kitaev_2021}. 
When expanding around $M=\infty$, it was conjectured that a transition into a glassy phase occurs at 
an exponentially small critical temperature $\log(T_{\rm g})\propto -\sqrt{M}$~\cite{georges_quantum_2001}, and 
a recent study has established that this phase displays full replica symmetry breaking (RSB)~\cite{christos_spin_2022}. 
A glass phase with similar features was found to occur 
in the transverse field Ising model~\cite{andreanov_long-range_2012,kiss_exact_2023} and the related 
multi-component quantum rotor models~\cite{YeSR93, read_landau_1995,tikhanovskaya_2023}.
In contrast, for bosonic $\mathrm{SU}(M)$ representations, a spin glass phase occurs even for $M=\infty$, albeit with one-step RSB~\cite{GeorgesPS2000,georges_quantum_2001}.

There is a fundamental difference between glasses displaying full RSB and one-step RSB. 
The energy landscape of the former features marginally stable local minima with a gapless spectrum of excitations, 
while the latter display a fully stable, gapped ground state, lying far below the manifold of marginally stable excited states that trap the dynamics. Marginal stability is key to the dynamics and the physics of avalanches in full RSB 
systems~\cite{muller_wyart_2015}. 
Which of these two scenarios applies to the physically relevant Heisenberg $\mathrm{SU}(2)$, $S=1/2$ spin glass 
is an open question. Exact diagonalization of small fully-connected systems is limited by finite size effects~\cite{arrachea_dynamical_2001,shackleton_quantum_2021}. Surprisingly, not much is known about the {\it classical } (large $S$) limit of the Heisenberg mean-field spin glass either, apart from the replica-symmetric analysis of Refs.~\cite{BrayMoore1981,Franz22}.  

The equilibrium quantum spin dynamics of the SY spin-liquid~\cite{sachdev_gapless_1993} is similar to that of a marginal Fermi liquid~\cite{varma_MFL}. It was realized early on~\cite{parcollet_non-fermi-liquid_1999} that this opens a new perspective 
on `strange' metals, culminating in recent models in which disorder and strong interactions conspire to prevent 
the emergence of quasiparticles and lead to $T$-linear resistivity down to the 
lowest temperatures~\cite{Patel2023, chowdhury_sachdev-ye-kitaev_2021}. It has recently been shown that such a behavior is found in the quantum critical region around the melting point of a metallic $\mathrm{SU}(2)$ Heisenberg spin glass~\cite{cha_linear_2020,dumitrescu_planckian_2021}. However, it is not known how the quantum dynamics change within the metallic spin glass phase itself. 

 In this Letter, we answer these open questions and present a solution of the
 fully connected spin-$1/2$ Heisenberg spin glass throughout its ordered phase.
 In the quantum case, the model reduces to an impurity problem for the spin dynamics
 coupled to Parisi's equations for the spin glass order.
 The Heisenberg glass has a full RSB solution, whose
 structure, however, differs crucially from an Ising glass. The
 insulating quantum glass displays the universal spin dynamics found in all
 solvable mean-field quantum glasses with full RSB in the limit $T \to 0$, but
 deviations from it persist to surprisingly low temperatures. 
 Upon doping, we find a metallic spin glass with unexpectedly slow spin dynamics, 
 close to the SY dynamics that dominate the quantum critical melting point of the spin glass.
 
%%%%%%%%%%%%%%%%%%%%%%%%%%%%%%% Section %%%%%%%%%%%%%%%%%%%%%%%%%%%%%%%
\paragraph{\textbf{Model and method}.} 
We consider $N$ spins $\S_i$ with all-to-all interactions, described by the Hamiltonian 
\begin{equation}
\mathcal{H} = - \sum_{i < j} J_{ij} \S_i \cdot \S_j.
\label{hamiltonian}
\end{equation}
The $\S_i$ are either classical Heisenberg spins -- vectors constrained to the
three-dimensional sphere of radius $S = 1/2$ -- or quantum $\rm SU(2)$ spins $S$ (with $\S^2 = S(S+1)$).  
We use $\hbar=k_B=1$ and denote by 
$\ell = 3$ the number of spin components. $J_{ij}$ are Gaussian random couplings
with zero mean and variance $J^2/N$. 
We obtain the equilibrium solution of this model using 
the replica method and Parisi's full RSB ansatz
~\cite{sk1975,bray_replica_1980,parisi_sequence_1980,duplantier_comment_1981,MPV_3,cr2002}, 
as detailed in the Supplementary Information~\cite{supp}. 

\rev{In brief, the} mean field model \rev{is} reduced to an effective single spin problem in a random frozen field $\h$ with distribution $\mathbb{P}(\h)$. \rev{In the classical case, it is governed by the Hamiltonian $H_{\rm loc}(\h) = - \h \cdot \S$}.
In the quantum case, the dynamics of this spin also depend on the self-averaging spin autocorrelation function
$\chi(\tau) = \langle \S(0) \S(\tau) \rangle - \langle \S \rangle^2$, via the action
\begin{equation}
\begin{split}
   \mathcal{S}_{\rm loc}(\h) = &\frac{J^2}{2} \int\!\!\!\! \int_0^{\beta} \d \tau \d \tau' \chi(\tau-\tau')
\S(\tau) \cdot \S(\tau')\\ 
&- \h \int_0^{\beta} \d \tau \, \S(\tau),
\end{split}
\label{Sloc}
\end{equation}
where $\tau$ is Matsubara imaginary time and $\beta = 1/T$ the inverse temperature. \rev{The glass phase is described by an order parameter $q(x)$ ($x \in [0,
1]$). $q(x)$ characterizes the distribution of phase space distances between local minima of the free energy landscape~\cite{MPV_3}. It determines}  
the local field distribution $\mathbb{P}(\h) \equiv
\mathbb{P}(x=1, \h)$, \rev{which} is found by solving Parisi's equations for the
magnetization $\mathbf{s}(x,\h)$ and frozen fields $\mathbb{P}(x, \h)$ 
\begin{subequations}
\label{parisiP}
\begin{align}
   \frac{\partial \mathbf{s}}{\partial x} &=  -\frac{J^2}{2} \frac{\d q}{\d x} \left( \del^2 \mathbf{s} + 2\beta x (\mathbf{s} \cdot \del) \mathbf{s} \right), \\ 
   \frac{\partial \mathbb{P}}{\partial x} &= \hphantom{-}\frac{J^2}{2} \frac{\d q}{\d x} \left( \del^2 \mathbb{P} - 2 \beta x\del(\mathbf{s} \cdot \mathbb{P}) \right),
 \\
\mathbf{s}(1,\h) %= \overline{\S}(\h)
&= \langle \S \rangle_{\rev{H_{\rm loc} (\h) |} \mathcal{S}_{\rm loc}(\h)}, 
\\ %\quad
   \mathbb{P}(0,\h) &= \delta(\h).
\end{align}
\end{subequations}
\rev{These equations are solved self-consistently, with}
\begin{equation}
q(x) = \frac{1}{\ell} \int \d \h \, \mathbb{P}(x,\h) \mathbf{s}(x, \h)^2,  
\label{qs2}
\end{equation}
and, in the quantum case,
\begin{equation}
\chi(\tau) = \frac{1}{\ell} \int  \d \h \, \mathbb{P} (\h) \,\langle \S(0) \S(\tau) \rangle_{\mathcal{S}_{\rm loc}(\h)} - q(1).  
\label{sc}
\end{equation}
Such self-consistency conditions are an example of extended dynamical mean-field theory (EDMFT)~\cite{sengupta_non-fermi-liquid_1995,smith_si_prb_2000,Chitra2000,andreanov_long-range_2012}.
\begin{figure}[ht]
%\centering
\includegraphics[scale=0.95]{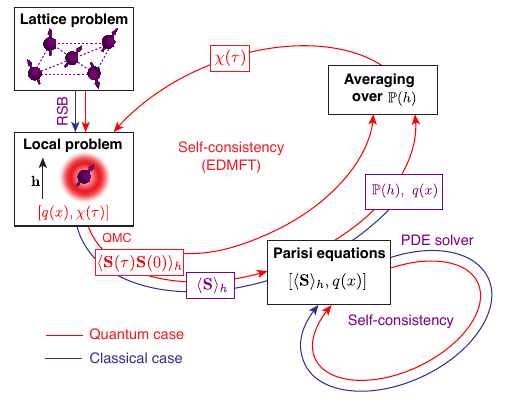}
\caption{
\label{figAlgoSolution}
Illustration of the algorithm to solve (\ref{Sloc},\ref{parisiP},\ref{qs2},\ref{sc}).
In the classical case, the local problem is solved only once, and $q(x)$ is obtained solving 
Parisi's partial differential equations.
In the quantum case, it is solved iteratively with CTQMC.
}
\end{figure}
The iterative procedure to solve them is summarized in Fig.~\ref{figAlgoSolution}.
A similar procedure has recently been implemented for mean field versions of transverse field Ising and quantum Coulomb glasses~\cite{kiss_exact_2023,Lovas2022}.
In contrast to the classical case, the single site quantum problem 
\eqref{Sloc} cannot be solved analytically. 
Its solution is obtained with a "CTSEG" continuous-time quantum Monte Carlo (CTQMC)
algorithm~\cite{otsuki_spin-boson_2013, supp}, without fermionic sign problem.
The Parisi equations are solved with a high precision numerical method
(error bars on $q(x)$ being of order $10^{-5}$), 
using their integral form~\cite{nemoto_numerical_1987} and filtering methods to
suppress numerical instabilities at low $T$~\cite{supp}.  
\begin{figure*}
\centering
\includegraphics[scale=0.95]{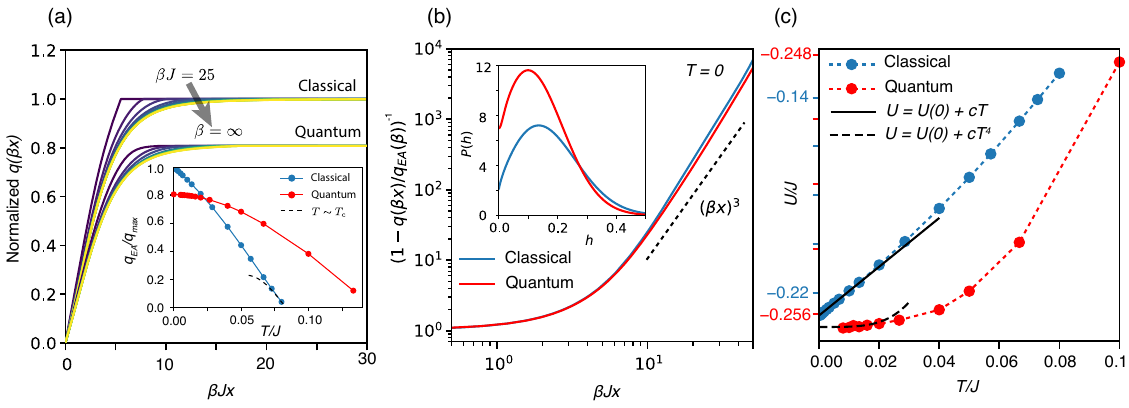}
\caption{
  \label{fig:sgphase}
   \textbf{Characteristics of the glass phase}. 
   (a) Overlap function $q(x)/q_{EA}(\beta)$, normalized by $q_{\rm
   EA}(T=0)/q_{\rm max}$, for a range of temperature values, from $\beta J = 25$ to $\beta = \infty$. 
   At low enough temperature, $q(x)/q_{EA}(\beta)$ depends only on $\beta x$. Inset: $q_{\rm EA}$ vs $T$.
   The dashed line is the result of \cite{gabay_symmetry_1982}, valid for $T\approx T_{\rm g}$.
   (b) $1-q(x)/q_{\rm EA}(T=0)$ at $T=0$,
   showing the power law approach to $q_{\rm EA}$ at large $\beta x$. 
   Inset: local field distribution $\mathbb{P}(h)$ at zero temperature.
%   (c) Breakpoint $x_c(T)$ vs $T$. 
%   $x_c(T)/T$ is a characteristic inverse energy scale governing the distribution of free energies of excited states.
%   It scales logarithmically with $T$ in both the classical and the quantum case. Dashed lines are guides to the eye. 
%   Inset: definition of the breakpoint $x_c$ ($q'(x>x_c)=0$).
(c) Internal energy as a function of temperature, with a fit for the classical case $U(T)- U(0) = cT$ ($ c = 1 \pm 0.1$), and for the quantum case $U(T)- U(0) = c T^4$.}
\end{figure*}

\paragraph{\textbf{Glass phase}.}
A spin glass phase appears below a critical temperature $T_{\rm g}$,
where the Edwards-Anderson order parameter $q_{\rm EA} \equiv q(x=1)$ turns on, 
as illustrated on Fig. \ref{fig:sgphase}a.
Our results for $T_{\rm g}$ are consistent with previous analytical work: $T_{\rm g} = JS^2/3$ ($\approx 0.08 J$ for $S^2=1/4$) in the classical
case~\cite{gabay_symmetry_1982} and $T_{\rm g} \approx J\langle \S^2\rangle/3\sqrt{3} 
\approx 0.15 J$ in the quantum case~\cite{bray_replica_1980}, with 
$q_{\rm EA}(T)$ following the prediction of \cite{gabay_symmetry_1982} close to $T_{\rm g}$ (Fig. \ref{fig:sgphase}a, inset).
Note that $T_{\rm g}$ is higher in the quantum case, since the quantum spins are larger ($S(S+1) > S^2$). 

For $T \ll T_{\rm g}$, the overlap function $q(x)$ 
obeys an \rev{approximate} scaling form $q(x, T) = q_{\rm EA}(T) f(x/T) + \rev{O(T/J)}$ (Fig. \ref{fig:sgphase}a \rev{and Fig. S13}).
\rev{The function $f$ is determined by solving} the Parisi equations (\ref{parisiP}ab) directly at $T=0$, \rev{upon changing to the natural variable $u = \beta x$}. 
In the quantum case, they require as boundary conditions the zero
temperature magnetizations, $\langle \S(\tau) \rangle_{\mathcal{S}_{\rm
loc}(\h)}$, which we approximate by our lowest temperature QMC
results (see \cite{supp} Fig. S5-S6). The results for $q(x)$ are
shown in Figs. \ref{fig:sgphase}a,b. The maximal possible overlap 
is
$q_{\rm max} =\langle \S^2\rangle/\ell$, corresponding to a product state without
fluctuations.
As expected, the classical glass approaches this value at $ T=0$,
while quantum fluctuations weakly reduce the order parameter to
$q_{\rm EA} \approx 0.81 q_{\rm max}$ (consistent with exact diagonalization in Ref.~\cite{arrachea_dynamical_2001} , while the fit of spin correlations in Ref.~\cite{shackleton_quantum_2021} likely underestimated the order).

Unlike in the Ising case, local stability does not require a pseudogap in the distribution of local fields. Instead $\mathbb{P}(\h=0)$ remains 
finite and $\mathbb{P}(h) \approx \P (0) + a h$ 
(cf. Fig. \ref{fig:sgphase}b), since
the probability of dangerously small fields is already suppressed by the phase
space factor $\sim h^{\ell-1}$~\cite{crisanti_analysis_2002}\cite{BrayMoore1981}. 
This in turn is related to the tail of the overlap function $q(x)$. Indeed, \eqref{qs2}
suggests that for $x$ close to 1, $q_{\rm EA} - q(x)$ counts the number of
spins that see frozen fields of order $T/x$, for which one may expect
$q_{\rm EA} - q(x) \sim \int_0^{T/x} \d h \, h^2 \P(h) \sim  \left({T}/{x}\right)^3 + O \left({T}/{x}\right)^4$, consistent with 
 an apparent power law: $\sim (T/x)^{\alpha}$.
Numerically we indeed find an apparent power law approach: $1 - q(x)/q_{\rm EA} \sim 1/(\beta x)^{\alpha}$ as $T \to 0$, with $\alpha$ slightly larger than 3. This contrasts with the Ising case~\cite{crisanti_analysis_2002, MPV_3,andreanov_long-range_2012}, 
where the linear pseudogap in $\P(h)$ leads to $\alpha=2$.
The lower density of small fields in the Heisenberg case suggests that
low-lying metastable states have higher energies. This
is consistent with the smaller value of the so-called breakpoint $x_c$, above which
$q(x)$ reaches a plateau
(\cite{supp}, Fig.~S12). $T/x_c(T)$ can be interpreted as the typical free energy
difference between the lowest metastable states~\cite{MPV_4}. While in the Ising
case that energy scales linearly with $T$ as $x_c(T \to 0) \approx 0.5$~\cite{crisanti_analysis_2002}, in Heisenberg glasses
it decreases much more slowly as  $T/x_c(T) \sim 1/\log(1/T)$,
corroborating the picture of a rougher energy landscape.
These differences will further show in the response to an increasing external
field, under which the ground state magnetization increases in
random discontinuous steps called `shocks'. Their size distribution $\rho(\Delta m)$ was numerically found to be very similar to that of field-triggered
out-of-equilibrium avalanches in the classical SK
model~\cite{Pazmandi99,le_doussal_avalanches_2010}, a phenomenon attributed to
the marginal stability of the full RSB landscape. 
The equilibrium $\rho(\Delta m)$ is governed by the asymptotic approach of $q(x)$ to $q_{\rm EA}$. If it scales as $(T/x)^{\alpha}$, then $\rho(\Delta m) \sim 1/(\Delta m)^{2/\alpha}$ for $N^{-1/2} \ll \Delta m \ll 1$ %(with an exponential cutoff at large $\Delta m$)
~\cite{le_doussal_avalanches_2010}. For the classical Ising glass with $\alpha=2$,  $\rho(\Delta m) \propto 1/(\Delta m)$ exhibits a broad spectrum of avalanches. The larger $\alpha\approx 3$ of Heisenberg glasses leads to $\rho(\Delta m) \propto 1/(\Delta m)^{2/3}$, with predominant large scale rearrangements -- as numerically observed in avalanches of XY ($\ell=2$) spins.~\cite{Sharma14}

\paragraph{\textbf{Specific heat}.}
Let us now consider the internal energy per spin.
In the classical case, it is given by~\cite{supp}:
\begin{equation}
U_{\rm Cl} = - \frac{  \beta J^2 S^4}{2\ell} \left( 1-  \frac{\ell^2}{S^4} \int_0^1 \d x \, q(x)^2 \right). 
\label{Ucl}
\end{equation}
In the quantum case, denoting $Q(\tau)=\chi(\tau)+q_{\rm EA}$ and $\overline{\tau} = \tau/\beta$, it reads
\begin{equation}
U_{\rm Q} = - \frac{  \ell \beta J^2}{2} \left( \int_0^1 \d \overline{\tau} \, Q(\overline{\tau})^2 -   \int_0^1 \d x \, q(x)^2 \right). 
\end{equation}
As shown in Fig. \ref{fig:sgphase}c, the classical and quantum internal
energies behave very differently as a function of temperature. For classical
vector spins, one expects a constant intra-state heat capacity $c=(\ell-1)/2$ as $T
\to 0$,  each degree of freedom contributing $1/2$ by the equipartition
theorem. 
A linear fit to our data yields $c \approx 1.0 \pm 0.1$, excluding sizeable inter-state
contributions, consistently with 
finite-size simulations~\cite{ching_monte_1976}. 
Quantum effects gap out the soft degrees of freedom and yield
a much weaker temperature dependence of the internal energy (Fig. \ref{fig:sgphase}c). \rev{A heat capacity proportional to $T^3$ was predicted for the SU$(N)$ quantum spin glass in the large $N$ limit, provided that marginally stable rather than equilibrium states are analyzed~\cite{schehr_low-temperature_2005}.}
Our data are compatible with $U(T) \propto T^4$, but only at the lowest temperatures $T/J\lesssim 0.02$, 
as finite temperature corrections are substantial. 
%This result is similar to several examples of quantum spin glasses solved by 1-step RSB where $c\propto T^3$ (provided that marginally stable  rather than equilibrium states are analyzed) \cite{schehr_low-temperature_2005}.

\begin{figure}
\centering
\includegraphics[scale=0.9]{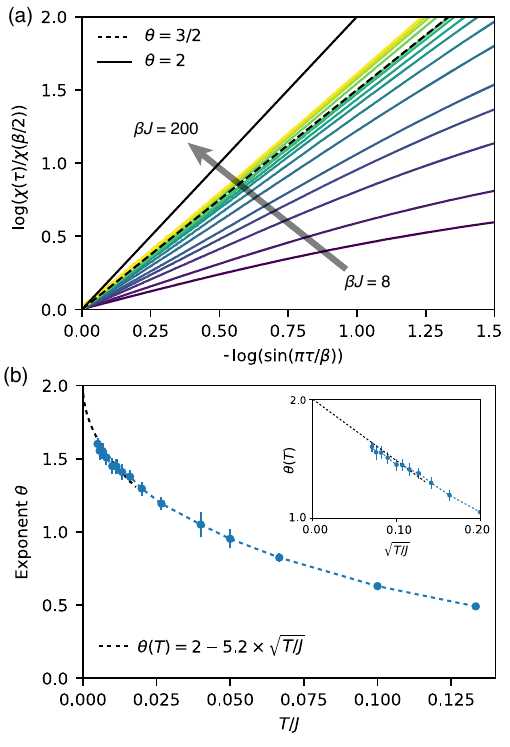}
\caption{
   \label{fig:susceptibility}
   \textbf{Spin susceptibility}. (a) Rescaled spin autocorrelation function in imaginary time, $\chi(\tau)/\chi(\beta/2)$, for a range of temperatures (from $\beta J = 8$ to $\beta J = 200$). At low temperature, it is well described by the conformal scaling form 
$\chi(\tau)/\chi(\beta/2)\approx \left[1/\sin(\pi \tau/\beta)\right]^{\theta}$, as shown by the linear behavior in the logarithmic plot. (b) Exponent $\theta$, obtained by fitting $Q(\tau)$ with the conformal scaling form, as a function of temperature. Inset: $\theta$ as a function of $\sqrt{T/J}$. The data is well described by $\theta (T) = 2 - 5.2 \times \sqrt{T/J}$ (black dashed line). 
}
\end{figure}

\paragraph{\textbf{Spin susceptibility}.} 

The spin-spin correlator $\chi(\tau)$ is shown in Fig. \ref{fig:susceptibility}a. For large $\tau$ 
it is well described by the conformal scaling form 
$\chi(\tau) \approx \chi(\beta/2) / \left[\sin(\pi \tau/\beta)\right]^{\theta}$~\cite{parcollet_non-fermi-liquid_1999}, 
implying that for $\omega \gtrsim T$, the dissipative part of the susceptibility at real frequencies $\chi''(\omega)$ scales as $\omega^{\theta - 1}$. 
The exponent $\theta$  has significant, slow $T$-dependence 
(Fig. \ref{fig:susceptibility}b). 
Using a Landau expansion, Ref.~\cite{read_landau_1995} predicted that $\theta(T=0)=2$,  ($\chi''(\omega)\propto\omega$). 
This value was also found in a $1/M$ expansion of the $\mathrm{SU}(M)$ quantum spin glass~\cite{christos_spin_2022} and 
actually holds for all solvable cases of marginally stable states found so far~\cite{cugliandolo_quantum_2022}. 
From the limited numerically accessible temperature range, it is hard to unambiguously conclude whether $\lim_{T \to 0}\theta(T) = 2$. 
However, Fig. \ref{fig:susceptibility}b suggests that this limit is indeed approached, albeit very slowly, as we can fit our data by $\theta(T) \approx 2 - 5.2 \times \sqrt{T/J}$. Interestingly, a similar behavior was found recently in the transverse-field Ising spin glass~\cite{kiss_exact_2023}. 

As discussed in Refs.~\cite{andreanov_long-range_2012,cugliandolo_quantum_2022}, the low-$T$ behavior $c\propto T^3$, $\chi''(\omega)\propto\omega$ ($\theta=2$) can be rationalized by approximating the eigenmodes of the Hessian describing the local curvature of the energy landscape as independent oscillators with spring constants $\lambda$ distributed as the eigenvalues of a random matrix $\rho(\lambda) \sim \sqrt{\lambda}$. For undamped oscillators of mass $M$, one has $M \omega^2 \sim \lambda$ with zero-point amplitudes  given by $\lambda\langle x^2 \rangle_{\omega} \sim \hbar \omega$, leading to $\chi''(\omega) \sim \rho(\omega) \langle x^2 \rangle_{\omega} \sim \sqrt{M}\omega$.  
In this  picture, an effective exponent $\theta < 2$ at finite $T$ may result from a $T$-dependent friction among modes. 
%It may also indicate that interactions between modes are significant at finite-$T$, but become irrelevant in the $T=0$ limit. 
%
Assuming $\lim_{T \to 0}\theta(T) = 2$, 
we have $A \equiv \lim_{\omega \to 0}\chi''(\omega) / \omega \approx \lim_{\beta \to \infty} \beta^2 \chi(\beta/2)/\pi$. We find $A \approx 3.5 J^2$ and thus
a significantly larger density of soft excitations than in the transverse
field Ising spin glass, where $A \sim 0.5 J^2$  (independent of the transverse
field) was reported~\cite{andreanov_long-range_2012,kiss_exact_2023}. 

\begin{figure}
\centering
\includegraphics[width=7.5cm]{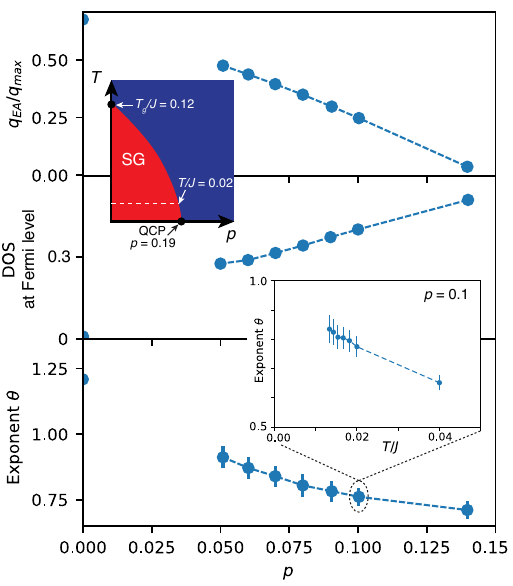}
\caption{
   \label{fig:metallicSG}
   \textbf{Doped case}.
Glass order parameter $q_{\rm EA}$, density of states at the Fermi level, and
scaling exponent $\theta$ of the spin-spin correlation function, as a function
of doping $p$ at $\beta J = 50$. \rev{Dashed lines are guides to the eye}. 
Upper inset: sketch of phase diagram. The spin glass (SG) melts at the quantum critical point (QCP).
Lower inset:  $\theta$ at fixed doping $p= 0.1$, as a function of temperature.
}
\end{figure}

\paragraph{\textbf{Metallic quantum spin glass}.}
In order to study the interplay between electrons and frozen spins arising from doping the spin glass, 
we use the Hamiltonian:
\begin{equation}
\label{Htj}
\mathcal{H} = - \sum_{ij,\sigma = \uparrow,\downarrow} t_{ij}c^{\dagger}_{i\sigma}c_{j \sigma} + U\sum_i n_{i \uparrow} n_{i \downarrow} - \sum_{i < j} J_{ij} \S_i \cdot \S_j, 
\end{equation}
where $c_{i\sigma}$ and $c^{\dagger}_{i\sigma}$ are the electronic annihilation and creation operators, 
$n_{i\sigma} = c^{\dagger}_{i\sigma} c_{i\sigma}$, $\S_i^a = c^\dagger_{i\sigma} \mathbf{\sigma}^a_{\sigma \sigma'} c_{i\sigma'}$  and $U$ is the on-site electron-electron interaction. 
The hopping amplitudes $t_{ij}$ are randomly distributed with variance $t^2/N$.  
We denote the doping by $p = \langle n_{\uparrow} + n_{\downarrow} \rangle - 1$.
This model has previously been solved in the paramagnetic
phase~\cite{dumitrescu_planckian_2021}. 
Here, we solve the spin glass phase; as in \cite{dumitrescu_planckian_2021}, we use $U = 4t$ and $J = 0.5t$. 

Below a critical doping $p_c$, a metallic spin glass appears, {\it i.e.} a phase with both a  non-zero spin glass order parameter $q_{\rm EA}$
and a non-zero density of states at zero frequency, as illustrated in Fig.~\ref{fig:metallicSG}.
This is compatible with the spin glass instability of the paramagnetic solution found in \cite{dumitrescu_planckian_2021}.
Previous studies of metallic quantum spin glasses have found a local spin susceptibility 
$\chi''(\omega) \propto \sqrt{\omega}$ corresponding to an exponent $\theta=3/2$ at $T = 0$ 
~\cite{sengupta_non-fermi-liquid_1995,sachdev_quantum_1995,cugliandolo_quantum_2022}. 
This can be interpreted as originating from Ohmic damping of the oscillators 
in the physical picture discussed above.
Intriguingly, we find that $\theta$ is smaller than unity 
for the whole range of (low) temperatures investigated.
Our data do not appear consistent with $\theta$ reaching $3/2$ at $T=0$ (possibly due to non-Ohmic damping by the non-Fermi-liquid metal) but 
may be consistent with $\theta$ reaching 1.
The latter corresponds to the quantum critical dynamics found at the critical point $p = p_c$, 
which might extend through a large part of the metallic spin-glass phase. 

In conclusion, we have solved the Heisenberg quantum spin glass for $\rm SU(2)$ spins.
We found that the energy landscape of vector spins differs significantly
from the Ising counterpart, resulting in different long time dynamics. Upon
decreasing $T$, the short time quantum dynamics slowly approach the super-universal
behavior of marginal mean field glasses, although featuring significantly
softer collective modes and a lower freezing temperature than comparable Ising glasses. The marginal Fermi liquid-type quantum dynamics anticipated from the $\mathrm{SU}(M\gg1)$ approach is essentially absent in the undoped insulating
limit of $SU(2)$ spins, but appears to be present in a wide window influenced
by the doping-induced quantum critical point in a metallic regime, that may be relevant for strongly correlated doped materials. 

\begin{acknowledgments}
The data and code associated with the paper are publicly available~\cite{github,zenodo}. We thank Philipp Dumitrescu, Subir Sachdev and Nils Wentzell for fruitful discussions. 
The Flatiron Institute is a division of the Simons Foundation. 
N.K. acknowledges support from a Humboldt fellowship. 
\end{acknowledgments}

\end{document}

\begin{titlepage}

\vspace*{2cm} \begin{center}
\large Supplemental Material for: \par\nobreak
\vspace{20pt}\hrule\vspace{10pt}
\huge\textbf{
Exact numerical solution of the fully connected classical and quantum Heisenberg spin glass} \par\nobreak
\vspace{10pt}\hrule\vspace{.6cm}
  \vspace{.5cm} 
\large Nikita Kavokine, Markus M\"uller, Antoine Georges and Olivier Parcollet  \\
\vspace{1cm}
-\\
\vspace{1cm}
\end{center}
\vspace{2cm}
\tableofcontents
\end{titlepage}

\section{Solution of the mean-field Heisenberg spin glass}

In this section, we describe the analytical procedure that reduces the mean-field (infinite dimensional) Heisenberg spin glass problem to a self-consistent single spin problem, amenable to an exact numerical solution. We make use of the replica method, as introduced by Sherrington and Kirkaptrick in the classical case~\cite{sherrington1975} and by Bray and Moore in the quantum case~\cite{bray_replica_1980}. We then follow Parisi's seminal replica symmetry breaking construction~\cite{parisi_sequence_1980, duplantier_comment_1981}, which was generalized to the quantum case of the transverse field Sherrington-Kirkpatrick model~\cite{andreanov_long-range_2012, kiss_exact_2023}, but not to the quantum Heisenberg spin glass.

\subsection{Model definition}

We consider an all-to-all interacting system of $N$ spins $\s_i$, described by the Hamiltonian 
\begin{equation}
H = - \sum_{i < j} J_{ij} \s_i \cdot \s_j.
\end{equation}
The $J_{ij}$ are independent random variables with distribution $\mathbb{G} (J_{ij}|J^2/N)$, where $\mathbb{G}(x |v)$ is the normalized Gaussian with zero mean and variance $v$. We will consider both a classical and a quantum version of the model. In the classical case, the $\s_i$ are vectors constrained to reside on the $\ell$-dimensional sphere of radius $S$. In the quantum case, the $\s_i$ are operators, forming a representation of spin $S$ of the group SU(2); we denote $\ell = 3$ the number of spin components.

\subsection{Replica trick}

Our aim is to compute $f$, the thermodynamic limit of the free energy per site at inverse temperature $\beta$. To this end, we introduce the quantity $f_n(N) \equiv (-1/\beta N) \log (\overline{Z^n })$, where the bar denotes averaging over the disorder. $Z^n$ is the partition function of $n$ replicas of the $N$-spin system, with the same realization of the disorder. Allowing $n$ to take non-integer values through analytical continuation, we find
\begin{equation}
\frac{f_n(N)}{n} = - \frac{1}{\beta n N} \log \left( \overline{ e^{n \log Z} } \right) \underset{n\to 0}{=} - \frac{1}{\beta  N} \overline{ \log Z }.
\label{nto0}
\end{equation}
Since the free energy is self-averaging in the thermodynamic limit, 
\begin{equation}
f = \lim_{N \to \infty} \lim_{n \to 0} \frac{f_n(N)}{n}. 
\end{equation}
We will assume that the order of the limits can be reversed, and first carry out the computation of $f_n$ in the thermodynamic limit. 

\subsubsection{Classical case}

We average over the random couplings the partition function of $n$ replicas of the spin system, labeled by the index $a$: 
\begin{align}
\overline{Z^n} &= \sum_{\{\s_i^a\}} \int_{-\infty}^{+\infty} \prod_{i < j} \d J_{ij} \, \mathbb{G} (J_{ij}|J^2/N) \exp \left[ \beta \sum_{a} \sum_{i < j} J_{ij} \s_i^a \cdot \s_j^a  \right] \\
&= \sum_{\{\s_i^a\}} \exp \left[ \frac{\beta^2 J^2}{2 N} \sum_{i < j} \sum_{a,b} (\s_i^a \cdot \s_j^a ) (\s_i^b \cdot \s_j^b ) \right] \\
&= \sum_{\{\s_i^a\}} \exp \left[ \frac{\beta^2 J^2}{4 N}\sum_{\alpha, \beta} \sum_{a,b}\left( \sum_i S_{i,\alpha}^a \cdot S_{i,\beta}^b  \right)^2 - n^2  \frac{\beta^2 J^2 N S^4}{4} \right].
\end{align}
Here, the indices $\alpha, \beta$ denote the spin components. We may drop in the following the term proportional to $n^2$, since it vanishes in the $n\to 0$ limit. We now perform a Hubbard-Stratonovitch transformation to decouple the sum over spins: 
\begin{equation}
\overline{Z^n} = \int \prod_{a , b, \alpha, \beta} \frac{\d Q^{\alpha \beta}_{ab}}{\sqrt{4\pi / ( N \beta^2 J^2)}} e^{-N \mathcal{F}[\{Q^{\alpha \beta}_{ab} \}]}, 
\label{NFcl}
\end{equation}
with 
\begin{equation}
\mathcal{F}[\{Q^{\alpha \beta}_{ab}  \}] =  \frac{\beta^2 J^2}{4} \sum_{a, b} \sum_{\alpha, \beta} (Q_{ab}^{\alpha \beta})^2 - \log \left[ \sum_{\{\s^a\}}  \exp \left( \frac{\beta^2 J^2}{2} \sum_{a,b} \sum_{\alpha, \beta} Q^{\alpha \beta}_{ab} \, S^a_{\alpha} \cdot S^b_{\beta} \right) \right].
\label{Fcl}
\end{equation}
In the thermodynamic limit, the integral in Eq.~\eqref{NFcl} is dominated by the saddle point $\{Q^{\alpha \beta \star}_{ab}  \}$: 
\begin{equation}
f_n(N) = (-1/\beta N) \log (\overline{Z^n }) \underset{N \to \infty}{\longrightarrow} \frac{1}{\beta} \mathcal{F}[\{Q_{ab}^{\alpha \beta \star} \}], 
\label{FforUcl}
\end{equation}
with $\partial \mathcal{F}/ \partial Q_{ab}^{\alpha \beta} |_{Q_{ab}^{\alpha \beta \star}} = 0, \forall (a, b, \alpha, \beta)$. The latter amounts to 
\begin{equation}
Q_{ab}^{\alpha \beta \star} = \langle S^a_{\alpha} \cdot S^b_{\beta} \rangle_{H_{\rm loc}[Q^{\star}}, ~~~~~ H_{\rm loc}[Q^{\star}] = - \frac{\beta J^2}{2}   \sum_{a,b}\sum_{\alpha,\beta} Q_{ab}^{\alpha \beta \star} S^a_{\alpha} \cdot S^b_{\beta}. 
\end{equation}

\subsubsection{Quantum case}

In the quantum case, we use an imaginary time path integral formalism to express the partition function: 
\begin{equation}
\overline{Z^n} = \sum_{\{\s_i^a\}} \int_{-\infty}^{+\infty} \prod_{i < j} \d J_{ij} \, \mathbb{G} (J_{ij}|J^2/N) \exp \left[ \beta \sum_{a} \sum_{i < j} J_{ij} \int_0^{1} \d \tau \, \s_i^a(\tau) \cdot \s_j^a(\tau)  \right],
\end{equation}
where the notation $\sum_{\s} [.]$ is now taken to mean the path integral $\int [D \s(\tau) ] e^{-\mathcal{S}_0 [\s (\tau)]} [.]$, where $\mathcal{S}_0 [\s (\tau)]$ is the imaginary time action describing an isolated spin, which we do not need to specify explicitly. We may then proceed as in the classical case: 
\begin{align}
\overline{Z^n} &= \sum_{\{\s_i^a\}} \exp \left[ \frac{\beta^2 J^2}{2 N} \int_0^1 \d \tau \d \tau' \sum_{i < j} \sum_{a,b} (\s_i^a(\tau) \cdot \s_j^a(\tau') ) (\s_i^b(\tau) \cdot \s_j^b(\tau') ) \right] \\
&= \sum_{\{\s_i^a\}} \exp \left[ \frac{\beta^2 J^2}{4 N}\int_0^1 \d \tau \d \tau' \sum_{\alpha, \beta} \sum_{a,b}\left( \sum_i S_{i,\alpha}^a(\tau) \cdot S_{i,\beta}^b(\tau')  \right)^2 - n^2  \frac{\beta^2 J^2 N}{4} S^2(S+1)^2  \right]
\end{align}
We may again drop the term proportional to $n^2$ as it vanishes in the $n \to 0$ limit. The Hubbard-Stratonovitch transformation yields
\begin{equation}
\overline{Z^n} = \int \prod_{a, b, \alpha, \beta} \int_0^1 \d \tau \d \tau' \, \frac{\d Q_{ab}^{\alpha, \beta}(\tau, \tau')}{\sqrt{4\pi / ( N \beta^2 J^2)}} e^{-N \mathcal{F}[\{Q^{\alpha \beta}_{ab}(\tau, \tau') \}]}, 
\end{equation}
with
\begin{equation}
\begin{split}
\mathcal{F}[\{Q_{ab} \}] =  &\frac{\beta^2 J^2}{4} \int_0^1 \d\tau \d\tau' \sum_{a, b} \sum_{\alpha, \beta} (Q_{ab}^{\alpha \beta}(\tau, \tau'))^2  \\ 
&- \log \left[ \sum_{\{\s^a\}}  \exp \left( \frac{\beta^2 J^2}{2} \int_0^1 \d \tau \d \tau' \sum_{a,b} \sum_{\alpha, \beta} Q^{\alpha \beta}_{ab} (\tau, \tau') \, S^a_{\alpha}(\tau) \cdot S^b_{\beta}(\tau') \right) \right].
\label{Fq}
\end{split}
\end{equation}
The saddle point approximation in the thermodynamic limit yields 
\begin{equation}
f_n(N) = (-1/\beta N) \log (\overline{Z^n }) \underset{N \to \infty}{\longrightarrow} \frac{1}{\beta} \mathcal{F}[\{Q_{ab}^{\alpha \beta \star}(\tau, \tau') \}], 
\label{FforUq}
\end{equation}
with $\partial \mathcal{F}/ \partial Q_{ab}^{\alpha \beta}(\tau,\tau') |_{Q_{ab}^{\alpha \beta \star}(\tau, \tau')} = 0, \forall (a, b, \alpha, \beta, \tau, \tau')$. The latter amounts to 
\begin{equation}
Q^{\alpha \beta \star}_{ab}(\tau, \tau') =  \langle \mathrm{T} \, S^a_{\alpha}(\tau) \cdot S^b_{\beta}(\tau') \rangle_{\mathcal{S}_{\rm loc}[Q^{\star}]},  
\end{equation}
\begin{equation}
\mathcal{S}_{\rm loc}[Q^{\star}] = \sum_a \mathcal{S}_0[\s^a(\tau)] - \frac{\beta^2 J^2}{2} \int_0^1 \d \tau \d \tau'  \sum_{a,b}\sum_{\alpha,\beta} Q_{ab}^{\alpha \beta \star}(\tau,\tau') S^a_{\alpha}(\tau) \cdot S^b_{\beta}(\tau'). 
\end{equation}

\subsubsection{Interpretation of the $Q_{ab}$}

We have reduced the initial lattice problem to a self-consistent local problem of $n$ coupled replicas in the limit $n \to 0$, which remains to be defined. We now establish the link between the spin-spin correlation functions of the local problem and those of the initial lattice problem. This will allow us to simplify some of the coefficients $Q_{ab}$ that define the local problem. For generality, we present here the reasoning in the quantum case. 

Let $\mathcal{S}$ be the action corresponding to the initial lattice problem. We supplement it with a source term: 
\begin{equation}
\mathcal{S} \mapsto \mathcal{S} + \beta^2 h \int_0^1 \d \tau \d \tau' \sum_i S_{i\alpha}(\tau) S_{i\beta} (\tau'). 
\end{equation}
Then, in the thermodynamic limit, 
\begin{equation}
\begin{split}
\int_0^1 \d \tau \d \tau' \, \overline{\langle \mathrm{T} S_{i\alpha}(\tau) S_{i\beta}(\tau') \rangle|_{\mathcal{S}}} &= \partial f/ \partial h|_{h= 0} \\
& = \lim_{n \to 0} \partial (f_n(N \to \infty)/ n)/\partial h |_{h = 0} \\
& = -\lim_{n \to 0} \frac{1}{n}\left. \frac{\partial}{\partial h}\right|_{h=0} \log \sum_{\{\s^a\}} e^{-\mathcal{S}_{\rm loc} [Q^{\star}] - h \int_0^1 \d \tau \d \tau' \sum_a S^a_{\alpha}(\tau) S^a_{\beta} (\tau')} \\
& = \lim_{n \to 0} \frac{1}{n} \sum_a \int_0^1 \d \tau \d \tau' \langle \mathrm{T} S^a_{\alpha}(\tau) S^a_{\beta}(\tau') \rangle |_{\mathcal{S}_{\rm loc} [Q^{\star}] } \\
& =  \lim_{n \to 0} \int_0^1 \d \tau \d \tau' \langle \mathrm{T} S^a_{\alpha}(\tau) S^a_{\beta}(\tau') \rangle |_{\mathcal{S}_{\rm loc} [Q^{\star}] }. 
\end{split}
\label{corresp1}
\end{equation}
Since the correlation function on the left hand side vanishes for two different spin components, the $Q_{aa}^{\alpha\beta\star}$ vanish if $\alpha\neq \beta$. We then set $Q(\tau - \tau') \equiv Q_{aa}^{\alpha\beta\star}(\tau, \tau')$. 
Consider now two copies of the initial system, governed by the action 
\begin{equation}
\mathcal{S}_{\rm II} = \mathcal{S}[\{\s_i^1\}] + \mathcal{S} [\{\s_i^2\}] 
\end{equation}
By applying the replica method to this action in the same way as above, we find the correspondence 
\begin{equation}
\overline{\langle \mathrm{T} S^1_{i\alpha}(\tau) S^2_{i\beta}(\tau') \rangle|_{\mathcal{S_{\rm II}}}} = \lim_{n \to 0, a \neq b}  \langle \mathrm{T} S^a_{\alpha}(\tau) S^b_{\beta}(\tau') \rangle |_{\mathcal{S}_{\rm loc} [Q^{\star}] } 
\end{equation}
As the copies 1 and 2 are independent, the right hand side of the above equation, and therefore the $Q_{ab}^{\alpha \beta\star}, a \neq b$, cannot depend on $\tau, \tau'$. Similarly, in the absence of quadrupolar order in the initial system, the $Q_{ab}^{\alpha \beta\star}$ vanish when $a \neq b$ and $\alpha \neq \beta$. The reasoning is analogous in the classical case. Thanks to the normalization of the classical spins, Eq.~\eqref{Fcl} may be further simplified by integrating over the $Q_{aa}^{\alpha \alpha}$ in Eq.~\eqref{NFcl}. 

Let us summarize our results so far.
\begin{framed}
In the classical case, the free energy per spin is 
\begin{equation}
f = -\frac{\beta J^2 S^4}{4\ell} + \lim_{n \to 0} \frac{1}{\beta n} \left[ \frac{\ell \beta^2 J^2}{2} \sum_{a < b} Q_{ab}^2 - \log Z_{\rm loc}^{\rm Cl} \right], 
\label{fQcl}
\end{equation}
with 
\begin{equation}
Z_{\rm loc}^{\rm Cl} =  \sum_{\{\s^a\}} e^{-\beta H_{\rm loc} }, ~~~~~ Q_{ab} = \frac{1}{\ell} \langle \s^a \cdot \s^b \rangle_{H_{\rm loc}}, ~~~~~ H_{\rm loc} = - \beta J^2   \sum_{a<b} Q_{ab} \s^a \cdot \s^b. 
\label{SCcl}
\end{equation}
\end{framed}
\begin{framed}
In the quantum case, 
\begin{equation}
f =   \frac{\ell\beta J^2}{4} \int_0^1 \d\tau  Q(\tau)^2 +  \lim_{n \to 0}  \frac{1}{\beta n} \left\{ \frac{\ell \beta^2 J^2}{2} \sum_{a < b} Q_{ab}^2
- \log Z_{\rm loc}^{\rm Q} \right\},
\label{fQq}
\end{equation}
with 
\begin{equation}
Z_{\rm loc}^{\rm Q} = \int \prod_a [D \s^a (\tau) ] e^{-\mathcal{S}_{\rm loc}} , ~ Q_{ab} = \frac{1}{\ell} \langle \mathrm{T} \, \s^a(\tau) \cdot \s^b(\tau') \rangle_{\mathcal{S}_{\rm loc}}, ~ Q(\tau) = \frac{1}{\ell} \langle \mathrm{T} \, \s^a(\tau) \cdot \s^a(0) \rangle_{\mathcal{S}_{\rm loc}},
\label{SCq}
\end{equation}
and
\begin{equation}
\begin{split}
\mathcal{S}_{\rm loc} = \sum_a \mathcal{S}_0[\s^a(\tau)] -  \frac{\beta^2 J^2}{2} \int_0^1 \d \tau \d \tau'   Q(\tau - \tau') \sum_{a} \s^a(\tau) \cdot \s^a(\tau')  \\ -
\beta^2 J^2  \sum_{a<b} Q_{ab} \int_0^1 \d \tau \s^a(\tau) \int_0^1 \d \tau' \s^b(\tau'). 
\end{split}
\end{equation}
\end{framed}
We now need to specify an ansatz for the matrix $Q_{ab}$ which allows for analytic continuation to non-integer matrix dimensions. 

\subsection{Replica symmetry breaking}

\subsubsection{Recursive construction}
We use Parisi's replica symmetry breaking (RSB) ansatz for the matrix $Q_{ab}$. The matrix $Q^{\mathrm{RSB}_k}$ at the $k^{\rm th}$ stage of RSB is defined by the choice of two sequences: $(q_p)_{p = 0..k}$ and $(m_p)_{p = 0..k}$, the latter being strictly decreasing, with $m_0 = n$. It is built "from the inside out" according to the following prescription: 
\begin{enumerate}
\item $Q^{\mathrm{RSB}_k}_k = (q_k-q_{k-1}) U_{m_k}$; 
\item $\forall p \in [0,k-1], Q^{\mathrm{RSB}_k}_p = \mathrm{diag}_{m_p/m_{p+1}}(Q^{\mathrm{RSB}_k}_{p+1}) + (q_{p}-q_{p-1}) U_{m_p}$;
\item $Q^{\mathrm{RSB}_k}_0 \equiv Q^{\mathrm{RSB}_k}$.
\end{enumerate}
Here $U_m$ is the matrix of size $m$ filled with ones of size, $\mathrm{diag}_r(A)$ is the matrix with $r$ times the block $A$ on the diagonal, and we set $q_{-1} = 0$. We note that this construction places $q_k$ instead of 0 on the diagonal, and the corresponding term needs to be subtracted in the end. In the classical case, the local Hamiltonian at the $k^{\rm th}$ stage of RSB is therefore
\begin{equation}
H_{\rm loc}^{\mathrm{RSB}_k} = n q_k \frac{\beta J^2 S^2}{2} - \frac{\beta J^2}{2} \sum_{ab} (Q^{\mathrm{RSB}_k})^{ab}\,  \s^a \cdot \s^b, 
\end{equation} 
which we may write as 
\begin{equation}
H_{\rm loc}^{\mathrm{RSB}_k} = \sum_a H_k [\s_a] - \frac{\beta J^2}{2} \sum_{ab} (Q^{\mathrm{RSB}_k})^{ab}\,  \s^a \cdot \s^b.
\end{equation} 
In the quantum case, we introduce the notation $\S \equiv \int_0^1 \d \tau \, \s (\tau)$. Then, the local action at the $k^{\rm th}$ stage of RSB is 
\begin{equation}
\mathcal{S}_{\rm loc}^{\mathrm{RSB}_k} = \sum_a \mathcal{S}_0[\s^a(\tau)] -  \int_0^1 \d \tau \d \tau'  \frac{\beta^2 J^2 (Q(\tau'-\tau)-q_k)}{2} \sum_a \mathbf{S}^a(\tau) \cdot \mathbf{S}^a(\tau') - \frac{\beta^2 J^2}{2}\sum_{ab} (Q^{\mathrm{RSB}_k})^{ab}\,  \S^a \cdot \S^b.
\end{equation}
For simplicity, we denote 
\begin{equation}
\mathcal{S}_{\rm loc}^{\mathrm{RSB}_k} = \sum_a \mathcal{S}_k [\s^a(\tau)] - \frac{\beta^2 J^2}{2}\sum_{ab} (Q^{\mathrm{RSB}_k})^{ab}\,  \S^a \cdot \S^b. 
\end{equation}

We now establish a recursion between the local partition functions $Z_{\rm loc}$ at steps $p$ and $p + 1$ of the recursive construction. We use the quantum notation, but the relation will be identical in the classical case. At step $p$ the local partition function under an external field $\h$ reads: 
\begin{equation}
Z_p^{\mathrm{RSB}_k}(\h) = \int \prod_a [D\s^a] \exp \left[ -\sum_a \mathcal{S}_k [\{\s^a\}] + \frac{\beta^2 J^2}{2}\sum_{ab} (Q^{\mathrm{RSB}_k}_p)^{ab}\,  \S^a \cdot \S^b + \beta \h \sum_a \S^a \right]. 
\end{equation}
We note that $\h$ is introduced only as an auxiliary term, and we will set $\h = 0$ in the end. In particular it does not represent an actual external field applied on the initial system: the matrix $Q_{ab}$ would then no longer be rotationally invariant. At the first step ($p=k$), we find 
\begin{equation}
\begin{split}
Z_k^{\mathrm{RSB}_k}(\h) &= \int \prod_a [D\s^a] \exp \left[ - \sum_{a=1}^{m_k} \mathcal{S}_k [\{\s^a\}] + \frac{\beta^2 J^2}{2}(q_k-q_{k-1}) \left(\sum_{a=1}^{m_k} \S^a \right)^2  + \beta \h \sum_{a=1}^{m_k} \S^a \right] \\
& = \int \d \h_k \, \mathbb{G}(\h_k|2J^2(q_k-q_{k-1})) \left(\int [D\s] e^{-  \mathcal{S}_k [\s]  + \beta (\h+\h_k)  \S } \right)^{m_k} \\
&\equiv \int \d \h_k \, \mathbb{G}(\h_k|2J^2(q_k-q_{k-1})) z_k(\h + \h_k)^{m_k}.
\end{split}
\end{equation}
At step $p$, we may consider separately the effect of the two terms in $Q^{\mathrm{RSB}_k}_p$: 
\begin{equation}
\begin{split}
Z_p^{\mathrm{RSB}_k}(\h) =  \int \prod_a [D\s^a] \exp \left[ - \sum_{a=1}^{m_p} \mathcal{S}_k [\{\s^a\}] + \frac{\beta^2J^2}{2}(q_p-q_{p-1}) \left(\sum_{a=1}^{m_p} \S^a \right)^2 \right. \\
  \left. + \frac{\beta^2 J^2}{2} \sum_{\alpha = 1}^{m_p/m_{p+1}} \sum_{a,b \in \alpha} (Q^{\mathrm{RSB}_k}_{p+1})^{ab} \S^a\cdot \S^b  + \beta \h \sum_{a=1}^{m_k} \S^a \right].
 \end{split}
\end{equation} 
Here, $\alpha$ labels the blocks of the matrix $Q^{\mathrm{RSB}_k}_p$. After a Hubbard-Stratonovitch transformation on the squared sum, we find the recurrence relation 
\begin{equation}
Z_p^{\mathrm{RSB}_k}(\h) = \int \d \h_p \, \mathbb{G}(\h_p|2J^2(q_p-q_{p-1})) \left[ Z^{\mathrm{RSB}_k}_{p+1}(\h + \h_p) \right]^{m_p/m_{p+1}}. 
\label{rec1}
\end{equation}

\subsubsection{Limit $n \to 0, k \to \infty$}
We now take the simultaneous limit $n \to 0$ and $k \to \infty$, keeping $m_k = 1$. Then, $m_p$ becomes a continuous variable $x$ between 0 and 1, and $q_p \to q(x)$, such that $m_p \to x$. $q(x)$ is the Parisi order parameter. The corresponding limit of the recurrence relation \eqref{rec1} may be taken if one uses the formal identity 
\begin{equation}
\left[ \exp \left( \frac{1}{2} v \boldsymbol{\nabla}^2\right) \cdot f \right] (\h) = \int \d \h' \mathbb{G}(\h'|v) f(\h+\h'), 
\end{equation}
that holds for any sufficiently regular function $f$. $Z_p^{\mathrm{RSB}_k}(\h)$ then becomes a continuous function $\zeta(x,\h)$, satisfying 
\begin{equation}
\zeta(x-\d x,\h) = \exp \left( \frac{1}{2} J^2 \d q(x) \del^2 \right) \zeta(x,\h)^{1-\d x/x}
\end{equation}
This may be recast into a partial differential equation: 
\begin{equation}
\frac{\partial \zeta}{\partial x} = -\frac{1}{2} J^2 \frac{\d q}{\d x} \del^2 \zeta + \frac{1}{x} \zeta \log \zeta,
\end{equation}
with boundary condition $\zeta(1,h) = z_{\infty}(\h)$ as defined below. At this point, a distinction needs to be made between the classical and the quantum case. In the classical case, the "single-replica" part of the local Hamiltonian $H_k$ does not depend on the spin. We may hence split it off in the partition function, so that applying Eq.~\eqref{rec1} at $p = 0$ yields 
\begin{equation}
\lim_{n \to 0} \frac{1}{n} \log Z_{\rm loc}^{\rm Cl} = -\frac{\beta^2 J^2 S^2}{2} q(1) + \int \d \h \, \mathbb{G}(\h | 2 J^2 q(0)) \lim_{x \to 0} \left[ \frac{1}{x} \log \zeta(x, \h) \right], 
\label{limZcl}
\end{equation}
and the boundary condition is explicitly
\begin{equation}
z_{\infty}^{\rm Cl}(\h) = \int \d \s e^{-\beta H_{\infty}(\h)} =  2 \pi \int_0^{\pi} \d \theta \sin \theta e^{-\beta h S \cos \theta } = \frac{2 \pi}{\beta h S}  (e^{\beta h S} - e^{-\beta h S}). 
\end{equation}
We may define $\phi(x, \h) \equiv (1/x) \log \zeta(x, \h)$, which satisfies 
\begin{equation}
\frac{\partial \phi}{\partial x} = -\frac{1}{2} J^2 \frac{\d q}{\d x}\left( \del^2 \phi + x (\del \phi)^2 \right).
\label{PDEphi}
\end{equation}
We also anticipate that $q(0) = 0$, so that the Gaussian in Eq.~\eqref{limZcl} becomes a $\delta$ function.
\begin{framed}
Then, introducing our results into Eq.~\eqref{fQcl} the free energy of the classical Heisenberg spin glass is given by 
\begin{equation}
f = - \frac{\beta J^2 S^4}{4 \ell} - \frac{\ell \beta J^2}{4} \int_0^1 \d x \, q(x)^2 + \frac{\beta J^2 S^2}{2} q(1) - \frac{1}{\beta} \phi(0, 0), 
\label{fcl}
\end{equation}
where $\phi(x,h)$ satisfies Eq.~\eqref{PDEphi} with $\phi(1, \h) = \log z^{\rm Cl}_{\infty} (\h)$.  
\end{framed}
We have used that, with the RSB ansatz~\cite{gabay_symmetry_1982}, 
\begin{equation}
\lim_{n \to 0} \sum_{a<b} Q_{ab}^2 = - \frac{1}{2} \int_0^1 \d x \, q(x)^2.
\end{equation}
This provides the solution of the classical spin glass problem given a function $q(x)$. 

In the quantum case, the single-replica part of the action cannot be simplified and we obtain 
\begin{equation}
\lim_{n \to 0} \frac{1}{n} \log Z_{\rm loc}^{\rm Q} = \int \d \h \, \mathbb{G}(\h | 2 J^2 q(0)) \lim_{x \to 0} \left[ \frac{1}{x} \log \zeta(x, \h) \right], 
\label{limZq}
\end{equation}
and 
\begin{equation}
z_{\infty}^{\rm Q}(\h) = \int [D \s] e^{- \mathcal{S}_{\infty}[\s (\tau), \h] }, 
\end{equation}
with 
\begin{equation}
\mathcal{S}_{\infty}(\h) = \mathcal{S}_0[\s (\tau)] - \int_0^1 \d \tau \d \tau' \frac{\beta^2 J^2 (Q(\tau - \tau') - q(1))}{2} \s(\tau) \cdot \s (\tau') + \beta \h \cdot \s
\end{equation}
\begin{framed}
Then, introducing our results into Eq.~\eqref{fQq} the free energy of the quantum Heisenberg spin glass is given by 
\begin{equation}
f =  \frac{\ell \beta J^2}{4} \left[ \int_0^1 \d\tau   Q(\tau)^2  - \int_0^1 \d x \, q(x)^2 \right] - \frac{1}{\beta} \phi(0, 0), 
\label{fq}
\end{equation}
where $\phi(x,h)$ satisfies Eq.~\eqref{PDEphi} with $\phi(1, \h) = \log z^{\rm Q}_{\infty} (\h)$.  
\end{framed}
This provides the solution of the quantum spin glass problem given $q(x)$ and $Q(\tau)$. 

\subsubsection{Local observables and two point functions}

We now determine the functions $q(x)$ and $Q(\tau)$ by applying the RSB ansatz to the self-consistency conditions in Eqs. \eqref{SCcl} and \eqref{SCq}. We go back to the recursive construction in 1.3.1 and consider the observable $R^a(\tau,\tau') =  \s^a(\tau) \s^a(\tau')$, which is local in replica space. 
For every $\tau, \tau'$, we define the quantity  $R^{\mathrm{RSB}_k}_p(\h)$, which corresponds to the average value of $R$ computed with the matrix $Q^{\mathrm{RSB}_k}$ built up to step $p$. $R^{\mathrm{RSB}_k}_p$ is only slightly different from $Z^{\mathrm{RSB}_k}_p$: indeed, at a given step, only one of the blocks is perturbed by the insertion of $R$, while the others are not. Therefore, one has the recurrence relation
\begin{equation}
R_p^{\mathrm{RSB}_k}(\h) = \int \d \h_p \, \mathbb{G}(\h_p|2J^2(q_p-q_{p-1})) \frac{R^{\mathrm{RSB}_k}_{p+1}(\h + \h_p)Z^{\mathrm{RSB}_k}_{p+1}(\h + \h_p)^{m_p/m_{p+1}}}{Z^{\mathrm{RSB}_k}_p(\h)}. 
\label{rec2}
\end{equation}
The situation is slightly different for an observable that is non-local in replica space, in particular the spin-spin correlator $\langle \S^a \S^b \rangle^{\mathrm{RSB}_k}$. When $k$ is sufficiently large, $a$ and $b$ are always in different blocks at $k$-RSB. If at step $p$ they are still in different blocks, then $\langle \S^a \S^b \rangle^{\mathrm{RSB}_k}_p = (\langle \S \rangle^{\mathrm{RSB}_k}_p)^2$. Let $p_0$ by the first step where $\S^a$ and $\S^b$ are in the same block. Then for $p < p_0$, the spin-spin correlator satisfies the same recurrence relation as a local observable: 
\begin{equation}
\langle \S^a \S^b \rangle^{\mathrm{RSB}_k}_p(\h) = \int \d \h_p \, \mathbb{G}(\h_p|2J^2(q_p-q_{p-1})) \frac{\langle \S^a \S^b \rangle^{\mathrm{RSB}_k}_{p+1}(\h + \h_p)Z^{\mathrm{RSB}_k}_{p+1}(\h + \h_p)^{m_p/m_{p+1}}}{Z^{\mathrm{RSB}_k}_p(\h)}. 
\label{rec3}
\end{equation}

In the limit $n \to 0, k \to \infty$, $R_p^{\mathrm{RSB}_k}(\h)$ becomes a function $\rho(x,\h)$, satisfying 
\begin{equation}
\zeta(x-\d x,\h)\rho(x-\d x,\h) = \exp \left( \frac{1}{2} J^2 \d q(x) \del^2 \right)\rho(x,\h) \zeta(x,\h)^{1-\d x/x}. 
\end{equation}
The corresponding PDE is 
\begin{equation}
\frac{\partial \rho}{\partial x} = -\frac{1}{2}J^2 \frac{\d q}{\d x} \left( \del^2\rho + 2 (\del \log \zeta) \cdot (\del \rho) \right), 
\label{PDErho}
\end{equation}
with the boundary condition $\rho(1,\h)_{\tau, \tau'} =  \langle\s(\tau) \s(\tau')\rangle_{\infty, \h}$, where $\langle \cdot \rangle_{\infty, \h}$ is the mean value with respect to $H_{\infty}(\h)$ or $\mathcal{S}_{\infty}(\h)$. If we now define $\beta \mathbf{s} (x,\h) = (1/x) \del \log \zeta$, we find that it satisfies the same equation as $\rho$, with the boundary condition $\mathbf{s}(1,\h) =  \langle \mathbf{S} \rangle_{\infty}$. Then $\rho$ satisfies 
\begin{equation}
\frac{\partial \rho}{\partial x} = -\frac{1}{2}J^2 \frac{\d q}{\d x} \left( \del^2\rho + 2 \beta x (\mathbf{s} \cdot \del) \rho \right) 
\label{PDErho2}
\end{equation}
Using Eq.~\eqref{rec2} with $p=0$,
\begin{equation}
 Q(\tau - \tau')  = \int \d \h \mathbb{G} (\h | 2J^2 q(0)) \rho(0,\h)_{\tau, \tau'} = \rho(0,0)_{\tau, \tau'}.
\label{RS}
\end{equation}
The spin-spin correlation function $\langle \mathbf{S}^a \mathbf{S}^b \rangle^{\mathrm{RSB}_k}_p$ becomes a continuous function $\nu(y,x,\h)$, where $y$ is the continuum analogue of the index $p_0$ defined above. For $y<x$, $\nu$ satisfies Eq.~\eqref{PDErho2}, with the boundary condition $\nu(y=x,x,\h) = \mathbf{s}^2(x,\h)$. The self-consistency conditions~\eqref{SCcl} and \eqref{SCq} become
\begin{equation}
q(x) = \frac{1}{\ell} \int \d \h \, \mathbb{G} (\h | 2J^2 q(0)) \nu(0,x,\h) = \frac{1}{\ell} \nu (0, x, 0). 
\end{equation}
This completes in principle our solution of the spin glass problem. One only needs to compute the average magnetization and the time correlation function in a single spin problem governed by $H_{\infty}$ or $\mathcal{S}_{\infty}$. This yields, through integration of Eq.~\eqref{PDErho2}, the functions $q(x)$ and $Q(\tau)$, which allow one in turn to compute the free energy, starting from the partition function associated with $H_{\infty}$ or $\mathcal{S}_{\infty}$. However, these results are formulated much more conveniently by introducing the probability distribution of the local field. 

\subsubsection{Probability distribution of the local field}

We introduce the probability distribution $\mathbb{P}(\h)$ of the local field acting on the effective impurity problem. Let us define the function $P(x,\h|x',\h')$ as the solution of the "backward" eq.~\eqref{PDErho2} with boundary condition $P(x',\h|x',\h') = \delta(\h-\h')$. 
For any $x' \geq x$, it has the Markovian property
\begin{equation}
P(x,\h|1,\h'') = \int \d \h' P(x,\h|x',\h')P(x',\h'|1,\h'').
\end{equation}
Differentiating with respect to $x'$ yields
\begin{equation}
0 = \int \d \h' \left[\partial_{x'} P(x,\h|x',\h')P(x',\h'|1,\h'') + P(x,\h|x',\h')\partial_{x'}P(x',\h'|1,\h'') \right], 
\end{equation}
and since $P$ satisfies~\eqref{PDErho2}, we obtain 
\begin{equation}
\begin{split}
\int \d \h' \, &P(x',\h'|1,\h'')\partial_{x'} P(x,\h|x',\h') =\\
& - \int \d \h' \, P(x,\h|x',\h') \frac{J^2}{2} \dot q(x') \left( \del_{\h'}^2 P(x',\h'|1,\h'') + 2 \beta x' (\mathbf{s}\cdot \del_{\h'}) P(x,\h'|1,\h'') \right). 
\end{split}
\end{equation}
Integrating by parts, 
\begin{equation}
\begin{split}
\int \d \h' \, &P(x',\h'|1,\h'')\partial_{x'} P(x,\h|x',\h') =\\
& \int \d \h' \, P(x',\h'|1,\h'') \frac{J^2}{2} \dot q(x') \left( \del_{\h'}^2 P(x,\h|x',\h') - 2 \beta x'\del_{\h'}(\mathbf{s}(x',\h') \cdot P(x,\h|x',\h') \right). 
\end{split}
\end{equation}
Since this is true for any $\h''$, $P(x,\h|x',\h')$ is found to satisfy the "forward" equation 
\begin{equation}
\frac{\partial P}{\partial x'} = \frac{J^2}{2} \dot q(x') \left( \del_{\h'}^2 P - 2 \beta x'\del_{\h'}(\mathbf{s} \cdot P \right)). 
\label{forward}
\end{equation}
We now rewrite eq.~\eqref{RS} as
\begin{equation}
\langle R \rangle_{\mathcal{S}} = \int \d \h' \d \h  \, \mathbb{G} (\h | 2J^2 q(0)) P(0,\h|1,\h') \langle R \rangle_{\mathcal{S}_1(\h')} \equiv \int \mathbb{P}(1,\h')\langle R \rangle_{\mathcal{S}_1(\h')}, 
\end{equation}
having defined 
\begin{equation}
\mathbb{P}(x',\h') = \int  \d \h  \, \mathbb{G} (\h | 2J^2 q(0)) P(0,\h|x',\h'). 
\end{equation}
$\mathbb{P}$ then satisfies eq.~\eqref{forward} with boundary condition  $\mathbb{P}(0,\h) = \mathbb{G} (\h | 2J^2 q(0)) = \delta(\h)$. Similarly, for the Parisi order parameter we find 
\begin{equation}
q(x) = \frac{1}{\ell} \int \d \h \, \mathbb{P}(x,\h) \mathbf{s}^2(x,\h)
\end{equation}

\subsubsection{Final formulation of the solution}

We are now in position to formulate the solution in its final form. 

\begin{framed}
We have reduced the solution of the fully connected Heisenberg spin glass problem to the solution of a single spin problem. It is governed in the classical case by the Hamiltonian $H_{\infty}(\h) = -\h \cdot \s$ and in the quantum case by the action 
\begin{equation}
\mathcal{S}_{\infty}(\h) = \mathcal{S}_0[\s (\tau)] - \int_0^1 \d \tau \d \tau' \frac{\beta^2 J^2 (Q(\tau - \tau') - q(1))}{2} \s(\tau) \cdot \s (\tau') - \beta \int_0^1 \h \cdot \s(\tau). 
\label{S1spin}
\end{equation}
The Parisi order parameter and local susceptibility are obtained as
\begin{equation}
q(x) = \frac{1}{\ell} \int \d \h\, \mathbb{P}(x,\h)  \mathbf{s}(x,\h)^2 ~~~~\mathrm{and} ~~~~~ Q(\tau) = \int \d \h \, \mathbb{P}(1, \h) \langle \mathrm{T} \s (\tau) \s(0) \rangle_{\infty, \h},
\label{main_result}
\end{equation}
where $\langle \cdot \rangle_{\infty,\h}$ denotes the mean value with respect to $H_{\infty}(\h)$ or $\mathcal{S}_{\infty}(\h)$. $\mathbb{P}(x,\h)$ and $\mathbf{s}(x,\h)$ are obtained by solving
\begin{align}
&\frac{\partial \mathbb{P}}{\partial x} = \frac{J^2}{2} \frac{\d q}{\d x} \left( \del^2 \mathbb{P} - 2 \beta x\del(\mathbf{s} \cdot \mathbb{P} )\right)~~ \text{with}~~ \mathbb{P}(0,\h) = \delta(\h);
\label{Pfinal} \\
& \frac{\partial \mathbf{s}}{\partial x} =  -\frac{J^2}{2} \frac{\d q}{\d x} \left( \del^2 \mathbf{s} + 2\beta x (\mathbf{s} \cdot \del) \mathbf{s} \right)~~ \text{with} ~~ \mathbf{s}(1,\h) = \langle \mathbf{S} \rangle_{\infty, \h}. 
\label{Sfinal}
\end{align}
Once $q(x)$ and $Q(\tau)$ are known, the free energy can be evaluated according to Eqs.~\eqref{fcl} and \eqref{fq}. 
\end{framed}
In practice, the local problem is solved trivially in the classical case: 
\begin{equation}
\langle S \rangle_{\infty, \h} = S \left( \frac{1}{\mathrm{tanh}(\beta S h)} - \frac{1}{\beta S h} \right), 
\end{equation}
and the self-consistency needs only to be established between Eqs. \eqref{main_result} through \eqref{Sfinal}. In the quantum case, the local problem is still a many-body problem, which can only be solved numerically. Moreover, the local action depends self-consistently on the observables $q(x)$ and $Q(\tau)$, so that convergence must be reached for two interconnected self-consistency loops (see Fig. 1 of the main text). 

\subsection{Internal energy}

The internal energy per spin $U/N$ can in principle be obtained by numerically differentiating the free energy with respect to temperature. However, it can be obtained in a more straightforward way by carrying out the differentiation analytically before replica symmetry breaking. Going back to Eqs.~\eqref{FforUcl} and \eqref{FforUq}, we compute the internal energy as 
\begin{equation}
\frac{U}{N} = - \frac{\partial (\beta f)}{\partial \beta} =  - \lim_{n \to 0} \frac{1}{n}\frac{ \partial \mathcal{F}[\{ Q_{ab}^{\star} \}]}{\partial \beta}. 
\end{equation}
Since the $Q_{ab}$ were introduced as integration variables, they may be arbitrarily rescaled. If we define $\tilde Q_{ab} = \beta^2 J^2 Q_{ab}$, the "complicated" terms in Eq.~\eqref{Fcl} and \eqref{Fq} no longer have an explicit $\beta$ dependence. We then obtain, in the classical case, 
\begin{equation}
\frac{\partial \mathcal{F}_{\rm Cl}}{\partial \beta} =  - \frac{n  \beta J^2 S^4}{2\ell} - \frac{\ell}{\beta^3 J^2} \sum_{a<b} \tilde Q_{ab}^2 =  - \frac{n \beta J^2 S^4}{2\ell} - \ell \beta J^2 \sum_{a<b} Q_{ab}^2.
\end{equation}
We then use the property 
\begin{equation}
\lim_{n \to 0} \sum_{a<b} Q_{ab}^2 = - \frac{1}{2} \int_0^1 \d x \, q(x)^2, 
\end{equation}
to obtain 
\begin{equation}
\frac{U_{\rm Cl}}{N} = - \frac{  \beta J^2 S^4}{2\ell} \left( 1-  \frac{\ell^2}{S^4} \int_0^1 \d x \, q(x)^2 \right). 
\end{equation}
Similarly, in the quantum case, 
\begin{equation}
\begin{split}
\frac{\partial \mathcal{F}_{\rm Q}}{\partial \beta} &= - \frac{  n \ell}{2\beta^3 J^2} \int_0^1 \d\tau \d\tau'  \tilde Q_{aa}(\tau, \tau')^2 - \frac{\ell}{\beta^3 J^2} \sum_{a<b} \tilde Q_{ab}^2 \\
&=  - \frac{n \ell \beta J^2}{2}\int_0^1 \d\tau \d\tau'  Q_{aa}(\tau, \tau')^2 - \ell \beta J^2 \sum_{a<b} Q_{ab}^2, 
\end{split}
\end{equation}
so that 
\begin{equation}
\frac{U_{\rm Q}}{N} = - \frac{  \ell \beta J^2}{2} \left( \int_0^1 \d \tau \, Q(\tau)^2 -   \int_0^1 \d x \, q(x)^2 \right). 
\end{equation}

\subsection{Doped quantum spin glass}

Our solution is readily generalized to a "doped" Heisenberg spin glass model. Physically, the randomly-interacting spins are now being carried by electrons, which can hop between the sites of a fully connected lattice and experience an on-site repulsion $U$. The corresponding Hamiltonian is 
\begin{equation}
H = H_0 + U\sum_i n_{i \uparrow} n_{i \downarrow} - \sum_{i < j} J_{ij} \S_i \cdot \S_j, 
\label{Htj}
\end{equation}
with $H_0 = - \sum_{ij,\sigma} (t_{ij} + \mu \delta_{ij}) c^{\dagger}_{i\sigma} c_{j \sigma}$. Here the $c_{i\sigma}$ and $c^{\dagger}_{i\sigma}$ are the electronic annihilation and creation operators, $\sigma = \uparrow,\downarrow$ labels the spin state, $n_{i\sigma} = c^{\dagger}_{i\sigma} c_{i\sigma}$ is the occupation number on site $i$, $U$ is the on-site electron-electron interaction, $\mu$ is the chemical potential and the $t_{ij}$ are hopping amplitudes, randomly distributed with variance $t^2/N$. The spins are defined in terms of the fermionic operators as $S_i^{\alpha} = (1/2) \sum_{\sigma, \sigma'} c_{i\sigma}^{\dagger} \boldsymbol{\sigma}^{\alpha}_{\sigma \sigma'} c_{i \sigma'}$, with $\boldsymbol{\sigma}^{\alpha}$ the Pauli matrices.  

The solution of this model proceeds along the same steps as above, and here we only outline the main differences. We represent the partition function as a path integral over the Grassman variables $(c_{i\sigma}^{\dagger}(\tau), c_{i\sigma}(\tau))$: 
\begin{equation}
Z = \int \prod_i [Dc_{i\sigma}^{\dagger}(\tau)][Dc_{i\sigma}(\tau)] e^{- \int_0^{\beta} \d \tau \left[\sum_{i,\sigma} c^{\dagger}_{i\sigma}(\tau) \partial_{\tau}  c_{i\sigma}(\tau) + H(\tau)\right]}.\end{equation}
Note that we now let the imaginary time run between 0 and $\beta$. We then compute $\overline{Z^n}$, with the disorder average now being carried out over both the couplings $J_{ij}$ and the hoppings $t_{ij}$. The Hubbard-Stratonovitch transformation now introduces integration variables $Q_{ab}$ and $\Delta_{ab}$, that decouple respectively the spin-spin interaction term and the hopping term. Upon performing the relevant simplifications (for instance, the $\Delta_{a \neq b}$ vanish), and dropping indices for simplicity, 
\begin{equation}
\overline{Z^n} \propto \int [DQ][D\Delta] \exp \left[ -N\left\{ \int_0^{\beta} \d \tau \d \tau' \left(\frac{J^2(Q^{ab})^2}{4 } + \frac{(\Delta^{a})^2}{t^2}\right) - \log \int [Dc]  e^{-\mathcal{S}_{\rm loc}[Q,\Delta]} \right\} \right],
\end{equation}
with 
\begin{equation}
\begin{split}
&\mathcal{S}_{\rm loc} = \int_0^{\beta} \d \tau \sum_{a\sigma}[ c_{\sigma}^{a\dagger}(\tau) (\partial_{\tau} - \mu) c_{\sigma}^{a}(\tau) + U n^a_{\uparrow} (\tau) n^a_{ \downarrow} (\tau) ] + \dots \\
& + \int_0^{\beta} \d\tau \d \tau' \sum_{a}\left[ \Delta (\tau'-\tau) \sum_{\sigma} c^{a\dagger}_{\sigma}(\tau') c^a_{\sigma}(\tau) - \frac{J^2 Q(\tau'-\tau)}{2} \mathbf{S}^a(\tau) \cdot \mathbf{S}^a(\tau') \right]- J^2\sum_{a<b} Q^{ab}  \mathbf{S}^a(\tau) \cdot \mathbf{S}^b(\tau').
\end{split}
\end{equation}
The saddle point equations read
\begin{align}
&\Delta(\tau) = t^2 G(\tau), ~~~ G(\tau) = - \langle \mathrm{T} c^a_{\sigma}(\tau) c^{\dagger a}_{\sigma} (0) \rangle; \\
&Q(\tau-\tau') = \frac{1}{\ell} \langle \mathrm{T} \mathbf{S}^a(\tau)\cdot \mathbf{S}^a (\tau') \rangle_{\mathcal{S}_{\rm loc}}; \\
&Q_{ab} =  \frac{1}{\ell} \langle \s^a(\tau) \cdot \s^b(\tau)  \rangle_{\mathcal{S}_{\rm loc}}.
\label{scq}
\end{align}

\begin{framed}
Upon replica symmetry breaking, the original lattice problem is reduced to a self-consistent impurity problem, governed by the action 
\begin{equation}
\begin{split}
&\mathcal{S}_{\infty}(\h) = \int_0^{\beta} \d \tau \sum_{\sigma}[ c_{\sigma}^{\dagger}(\tau) (\partial_{\tau} - \mu) c_{\sigma}(\tau) + U n_{\uparrow} (\tau) n_{ \downarrow} (\tau) ] - \int_0^{\beta} \d \tau \h \, \cdot\mathbf{S}(\tau) \\
& + \int_0^{\beta} \d\tau \d \tau' \left[ \Delta (\tau'-\tau) \sum_{\sigma} c^{\dagger}_{\sigma}(\tau') c_{\sigma}(\tau) - \frac{J^2 (Q(\tau'-\tau)-q(1))}{2} \mathbf{S}(\tau) \cdot \mathbf{S}(\tau') \right]. \\
\end{split}
\label{S1}
\end{equation}
The self-consistency conditions read 
\begin{equation}
q(x) = \frac{1}{\ell} \int \d \h\, \mathbb{P}(x,\h)  \mathbf{s}(x,\h)^2, ~~~~~ Q(\tau) = \int \d \h \, \mathbb{P}(1, \h) \langle \mathrm{T} \s (\tau) \s(0) \rangle_{\infty, \h},
\end{equation}
and 
\begin{equation}
\Delta(\tau) = t^2 \int \d \h \, \mathbb{P}(x, \h) \frac{1}{2} \sum_{\sigma} G_{\sigma}(\tau, \h), ~~~\mathrm{with} ~~~ G_{\sigma}(\tau) = - \langle \mathrm{T} c_{\sigma}(\tau) c^{\dagger }_{\sigma} (0) \rangle_{\infty, \h}. 
\label{sc_G}
\end{equation}
$\mathbb{P}(x,\h)$ and $\mathbf{s}(x,\h)$ are obtained by solving Eqs. \eqref{Pfinal} and \eqref{Sfinal}. 
\end{framed}

We note that we have used in Eq.~\eqref{sc_G} the spin-symmetrized Green's function  $\overline{G}(\tau) = (G_{\uparrow}(\tau) + G_{\downarrow}(\tau) ) / 2$. In the absence of an external magnetic field, it is equal to the Green's function of either spin. In the presence of a field $\h$, $G_{\uparrow}$ and $G_{\downarrow}$ depend on both the magnitude and direction of the $\h$, but $\overline{G}$ depends only on $|| \h ||$. 

Indeed, let us apply a unitary transformation $U$ to the basis of spin states $\{ | \sigma \rangle \}$. This defines new basis vectors $\{ | \sigma' \rangle \}$. We may define creation and annihilation operators for spin up and spin down electrons in this new basis. They are given by 
\begin{equation}
c^{\dagger}_{\sigma'} = \sum_{\sigma = \uparrow,\downarrow} U_{\sigma'\sigma} c^{\dagger}_{\sigma} ~~~~~~~~~~~\text{and}~~~~~~~~~~~c_{\sigma'} = \sum_{\sigma = \uparrow,\downarrow} U^{*}_{\sigma' \sigma} c_{\sigma} 
\end{equation}
The spin up Green's function in the new basis is
\begin{align}
G'_{\uparrow}(\tau) &= -\langle (U^{*}_{\uparrow \uparrow}  c_{\uparrow}(\tau) + U_{\uparrow \downarrow}^* c_{\downarrow} (\tau) ) (U_{\uparrow \uparrow}  c^{\dagger}_{\uparrow}(0) + U_{\uparrow \downarrow} c^{\dagger}_{\downarrow} (0) ) \rangle \\
& = |U_{\uparrow \uparrow} |^2 G_{\uparrow}(\tau) + |U_{\uparrow \downarrow} |^2 G_{\downarrow} (\tau) - U^{*}_{\uparrow \uparrow} U_{\uparrow \downarrow} \langle c_{\uparrow}(\tau) c^{\dagger}_{\downarrow}(0) \rangle -  U_{\uparrow \uparrow} U_{\uparrow \downarrow}^* \langle c_{\downarrow}(\tau) c^{\dagger}_{\uparrow}(0) \rangle
\end{align}
Similarly, 
\begin{align}
G'_{\downarrow}(\tau) &= -\langle (U^{*}_{\downarrow \uparrow}  c_{\uparrow}(\tau) + U_{\downarrow \downarrow}^* c_{\downarrow} (\tau) ) (U_{\downarrow \uparrow}  c^{\dagger}_{\uparrow}(0) + U_{\downarrow \downarrow} c^{\dagger}_{\downarrow} (0) ) \rangle \\
& = |U_{ \downarrow \uparrow} |^2 G_{\uparrow}(\tau) + |U_{\downarrow \downarrow} |^2 G_{\downarrow} (\tau) - U^{*}_{\downarrow \uparrow} U_{\downarrow \downarrow} \langle c_{\uparrow}(\tau) c^{\dagger}_{\downarrow}(0) \rangle -  U_{\downarrow \uparrow} U_{\downarrow \downarrow}^* \langle c_{\downarrow}(\tau) c^{\dagger}_{\uparrow}(0) \rangle
\end{align}
Now, the unitarity of $U$ imposes that $|U_{\uparrow \uparrow}|^2 + |U_{\uparrow \downarrow} |^2 = 1$ and $U_{\uparrow \uparrow} U_{\uparrow \downarrow}^* + U_{\downarrow \uparrow} U_{\downarrow \downarrow}^* = 0$, so that 
\begin{equation}
G'_{\uparrow}(\tau) + G'_{\downarrow}(\tau) = G_{\uparrow}(\tau) + G_{\downarrow}(\tau). 
\end{equation}
The spin-symmetrized Green's function is thus rotationally-invariant even under an external magnetic field. 

\section{Numerical procedures}

\subsection{Solution of the Parisi equations}
\subsubsection{Integral equation formulation}

The Parisi equations Eqs. \eqref{Pfinal} and \eqref{Sfinal} are non-linear partial differential equations that need to be solved numerically. We introduce the Green's function $G(x,\h|x',\h')$ of their linear part, which satisfies
\begin{equation}
\left\{
\begin{array}{lr}
\dfrac{\partial G}{\partial x} = \dfrac{J^2}{2} \dfrac{\d q}{\d x} \del^2 G & \text{for}~x\geq x' \\
& \\
\dfrac{\partial G}{\partial x} = -\dfrac{J^2}{2} \dfrac{\d q}{\d x} \del^2 G & \text{for}~x \leq x'
\end{array}\right. .
\end{equation}
with $G(x',\h|x',\h') = \delta(\h - \h')$. $G$ is simply a Gaussian: 
\begin{equation}
G(x,\h|x',\h') = \frac{1}{[2\pi J^2 |q(x) - q(x')|]^{3/2}} \exp \left( {-\frac{(\h-\h')^2}{2 J^2 |q(x) - q(x')|}} \right).
\end{equation}
If we now treat the non-linear terms as source terms, Eqs.~\eqref{Pfinal} and~\eqref{Sfinal} can be formally integrated according to 
\begin{equation}
\mathbb{P}(x,\h) = \frac{1}{(2\pi J^2 q(0))^{3/2}} e^{-h^2/2 J^2 q(0)} - \beta J^2 \int_0^x \d x' \, x' \dot q(x') \int \d \h' \, G(x,\h|x',\h') \del( \mathbf{s} \cdot \mathbb{P})|_{x',\h'}
\label{intP}
\end{equation}
and
\begin{equation}
\mathbf{s}(x,\h) = \int \d \h' G(x,\h|1,\h') \langle \mathbf{S} \rangle_{\infty, \h'} + \beta J^2 \int_x^1 \d x' \, x' \dot q(x') \int \d \h' \, G(x,\h|x',\h') (\mathbf{s} \cdot \del) \mathbf{s}|_{x',\h'}
\label{intS}
\end{equation}
We now use the fact that $\mathbf{s}(x,\h)$ is always in the direction of $\h$: $\mathbf{s}(x,\h) = s(x,h) \hat \h $. Then, 
\begin{equation}
\del( \mathbf{s} \cdot \mathbb{P})|_{x',\h'} = \frac{1}{h'^2} \frac{\partial(h'^2s \mathbb{P})}{\partial h'} 
\end{equation}
and 
\begin{equation} 
(\mathbf{s} \cdot \del) \mathbf{s}|_{x',\h'} = s \frac{\partial s}{\partial h'} \hat \h'. 
 \end{equation}
 Hence, we may introduce angle-integrated Green's functions: 
 \begin{equation}
 \begin{split}
 \overline{G} (x,h&|x',h') = (h')^2 \int \d \hat \h' \, G(x,\h|x',\h')  \\
 & =  \frac{1}{[2\pi J^2 |q(x) - q(x')|]^{1/2}} \frac{hh'}{h^2} \left[ \exp \left( {-\frac{(h-h')^2}{2 J^2 |q(x) - q(x')|}} \right) - \exp \left( {-\frac{(h+h')^2}{2 J^2 |q(x) - q(x')|}} \right)\right]; 
 \end{split}
 \end{equation}
  \begin{equation}
 \begin{split}
 \overline{G}_s (x,h&|x',h') = (h')^2 \int \d \hat \h' \, G(x,\h|x',\h') (\hat \h' \cdot \hat \h) \\
 & =  \frac{1}{[2\pi J^2 |q(x) - q(x')|]^{1/2}} \frac{1}{h^2} \left[ (J^2 | q(x) - q(x')| + hh') \exp \left( {-\frac{(h+h')^2}{2 J^2 |q(x) - q(x')|}} \right) \right. \\
& \left. + (hh' - J^2|q(x)-q(x')|)\exp \left( {-\frac{(h-h')^2}{2 J^2 |q(x) - q(x')|}} \right)\right]. 
 \end{split}
 \end{equation}
Eqs.~\eqref{intP} and \eqref{intS} then reduce to one-dimensional integral equations: 
 \begin{equation}
\mathbb{P}(x,h) = \frac{1}{(2\pi J^2 q(x))^{3/2}} e^{-h^2/2 J^2 q(x)} - \beta J^2 \int_0^x \d x' \, x' \dot q(x') \int \d h' \, \overline{G}(x,h|x',h') \frac{1}{h'^2} \frac{\partial(h'^2s \mathbb{P})}{\partial h'} 
\label{intP1D}
\end{equation}
and
 \begin{equation}
s(x,h) = \int \d h' \, \overline{G}_s(x,h|1,h') \langle S \rangle_{\infty, \h'} + \beta J^2 \int_x^1 \d x' \, x' \dot q(x') \int \d h' \, \overline{G}_s(x,h|x',h') \left. s \frac{\partial s}{\partial h'}\right|_{x',h'}.
\label{intS1D}
\end{equation}

\subsubsection{Numerical procedure}

Our solver code for Eqs.~\eqref{intP1D} and \eqref{intS1D}, written in Python, is available on Zenodo~\cite{zenodo}. 
The variable $x$ is discretized, and Eqs.~\eqref{intP1D} and \eqref{intS1D} are solved on successive $x$ points. We typically use $128$ to $256$ $x$ points in the region where $q(x)$ is non-constant. When $q(x)$ and $q(x')$ are very close, the Green's functions $\overline{G}(x,h|x',h')$ are strongly peaked around $h = h'$. However, the result of the convolution varies much more slowly as a function of $h$. Therefore, we sample $h'$ on a very fine grid (typically 20000 points) and $h$ on a much coarser Chebyshev grid (typically 256 points), and then use barycentric interpolation to obtain the result of the convolution on the fine grid for the next integration step. 

When the non-linear term is large (typically, when $q(x)$ is large), the numerical integration procedure is unstable. Indeed, at step $n$, the non-linear term contains a contribution from the $n^{\rm th}$ numerical derivative of the initial condition, which develops an oscillatory instability with a frequency on the order of the Chebyshev grid spacing. As long as the grid is much finer than the structure in $\mathbb{P}(x,h)$ and $s(x,h)$, the spurious oscillations may be filtered out. We use Savitzky-Golay filtering, which amounts to splitting the $h$ interval into panels, and fitting $\mathbb{P}(x,h)$ or $s(x,h)$ with a second degree polynomial in $h$ on each of the panels. The panel size is determined adaptively, as a fraction of the value of $h$ where the non-linear term has a maximum. With this procedure, and given our level of discretization, the filtered-out part is $\lesssim 10^{-5}$ of the remainder at every $x$ step, which roughly sets the precision level of our solver. We further check that the normalization $\int_0^\infty 4 \pi h^2 \mathbb{P}(x,h) \d h = 1$ and the sum rule~\cite{crisanti_analysis_2002}
\begin{equation}
1 - \frac{\ell}{S^2} \int_0^1 \d x \, q(x) = T
\label{sum_rule}
\end{equation}
in the classical case are satisfied to a precision better than $10^{-4}$ after self-consistency has been reached. 

\rev{
\subsubsection{Solution at $T = 0$}

Upon the change of variable $u = \beta x$, the Parisi equations~\eqref{Pfinal} and \eqref{Sfinal} become
\begin{align}
&\frac{\partial \mathbb{P}}{\partial u} = \frac{J^2}{2} \frac{\d q}{\d u} \left( \del^2 \mathbb{P} - 2 u \del(\mathbf{s} \cdot \mathbb{P} )\right)~~ \text{with}~~ \mathbb{P}(0,\h) = \delta(\h); \\
& \frac{\partial \mathbf{s}}{\partial u} =  -\frac{J^2}{2} \frac{\d q}{\d u} \left( \del^2 \mathbf{s} + 2u (\mathbf{s} \cdot \del) \mathbf{s} \right)~~ \text{with} ~~ \mathbf{s}(u  = \beta,\h) = \langle \mathbf{S} \rangle_{\infty, \h}. 
\end{align}

In this form, they may be solved directly in the limit $\beta \to \infty$ ($T = 0$). We then impose the boundary condition for $\mathbf{s}$ at a finite $u_{\rm max}$, ensuring that the value of $u_{\rm max}$ does not affect the solution in the relevant range of $u$. To obtain the data in Fig. 2b, we used $u_{\rm max} = 100/J$. 
}

\subsection{Solution of the quantum impurity problem}

We solve the quantum impurity problem defined by the action in Eq.~\eqref{S1}, and its restriction in Eq.~\eqref{S1spin}, by using a continuous-time Monte Carlo procedure based on a hybridization expansion in the segment picture (CTSEG algorithm) \cite{otsuki_spin-boson_2013, ayral}. 
Our quantum Monte Carlo code, based on the TRIQS library, is available on GitHub~\cite{github}. 
Briefly, the algorithm is based on the double expansion of the hybridization ($\Delta$) and spin-spin interaction ($Q$) terms. A configuration is defined by the order of the expansion and by the choice of points in imaginary time. Several possible updates or "moves" are defined for the configuration, which are accepted or rejected according to the Metropolis prescription. Further details of our implementation will be described elsewhere. 

\subsection{General solution procedure}

\subsubsection{Classical case}
In the classical case, the magnetization $\langle \s \rangle_{\infty, \h}$ is known analytically and does not depend on the spin glass parameters. Therefore, we start from an initial guess for $q(x)$ and iteratively solve Eqs.~\eqref{Sfinal} and \eqref{Pfinal} until convergence. As self-consistency is approached, $q(x)$ develops a singularity: $q(x) = q(1)$ for $x > x_c$. Yet, our numerical solution cannot produce a singularity in the mathematical sense. Two procedures can be envisioned for finding the breakpoint $x_c$. 
\begin{enumerate}
\item We may apply a threshold to $\d q /\d x$. Concretely, once a new $q(x)$ has been determined from Eq.~\eqref{main_result}, we enforce that if $q(1) - q(x) < \epsilon$, $q(x) = q(1)$, with typically $\epsilon = 10^{-5}$, on the order of our solver's accuracy. This will result in the breakpoint moving as the solution is iterated, and eventually converging. 
\item We may iterate the equations until convergence for a range of fixed values of $x_c$, and then choose the $x_c$ for which the sum rule in Eq.~\eqref{sum_rule} is most accurately satisfied. 
\end{enumerate}
We found that procedure 1 converges in a reasonable number of iterations at low temperatures $T/J \leq 0.04$, while procedure 2 is required for higher temperatures. At zero temperature, since $x_c / T$ logarithmically diverges as $T \to 0$, there is no breakpoint in the finite interval of $x/T$ in which we solve the equations. 

\subsubsection{Quantum case}
In the quantum case, the magnetization $\langle \s \rangle_{\infty, \h}$ is obtained by the solving the action in Eq.~\eqref{S1} or \eqref{S1spin} with quantum Monte Carlo for a range of values of $h$. Furthermore, the action depends on the spin glass parameters and must be updated until self-consistency is reached. We start from an initial guess for $q(x)$ and $Q(\tau)$ (and $\Delta(\tau)$ in the doped case), and determine $\langle \s \rangle_{\infty, \h}$ (as well as the spin-spin correlation function, and possibly the electron Green's function) on the impurity, for typically 30 values of $h$ sampled on a Chebyshev grid. The observables at arbitrary values of $h$ are then inferred by barycentric interpolation. Then, given $\langle \s \rangle_{\infty, \h}$, Eqs.~\eqref{Sfinal} and \eqref{Pfinal} are iterated $N$ times. An arbitrarily large value of $N$ is not necessarily best for rapid convergence of the whole self-consistent procedure; we found empirically that $N = 5$ is adequate. Procedure 1 (defined above) was sufficient to obtain reasonable convergence of the breakpoint. Once $q(x)$ and $\mathbb{P}(h)$ are obtained, we determine the $Q(\tau)$ and $\Delta(\tau)$ that define a new impurity action, and repeat the procedure until convergence of $q(x)$, $Q(\tau)$ and $\Delta(\tau)$. 

The Parisi equations \eqref{Sfinal} and \eqref{Pfinal} can be solved directly at zero temperature, but not the quantum impurity problem. However, the input for the Parisi equations is only the magnetization, which we find to be almost converged with temperature for $\beta J > 100$ (Fig. S5). We use the magnetization obtained at the lowest accessible temperature $\beta J = 200$ as an input for the zero-temperature Parisi equations. 

\newpage

\section{Supplementary figures}
\begin{figure}[hb]
\centering
\includegraphics[scale=0.65]{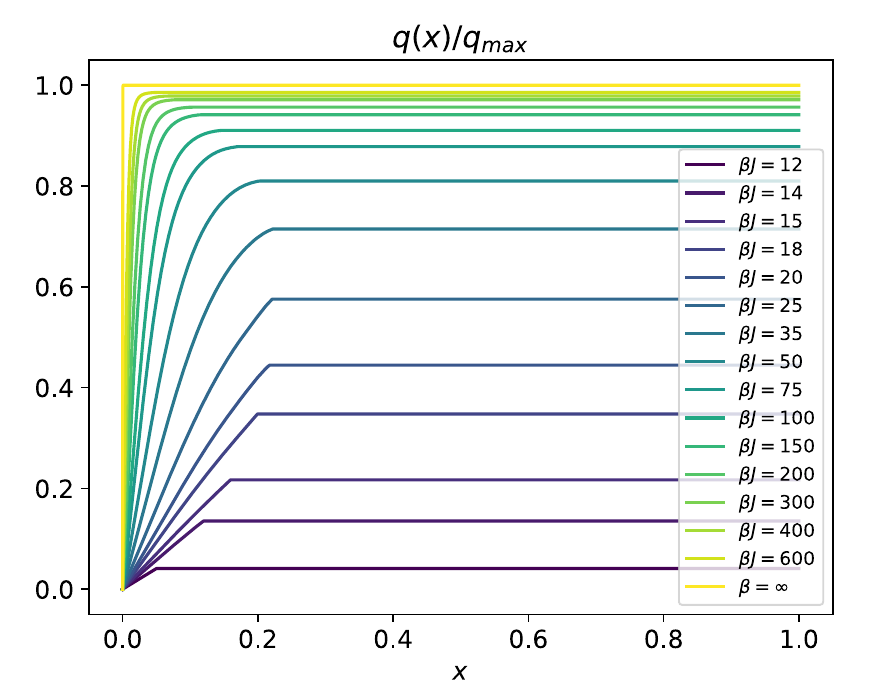}
\caption{Parisi order parameter $q(x)$ for the classical Heisenberg spin glass, for a range of temperature values.}
\end{figure}

\begin{figure}[ht]
\centering
\includegraphics[scale=0.65]{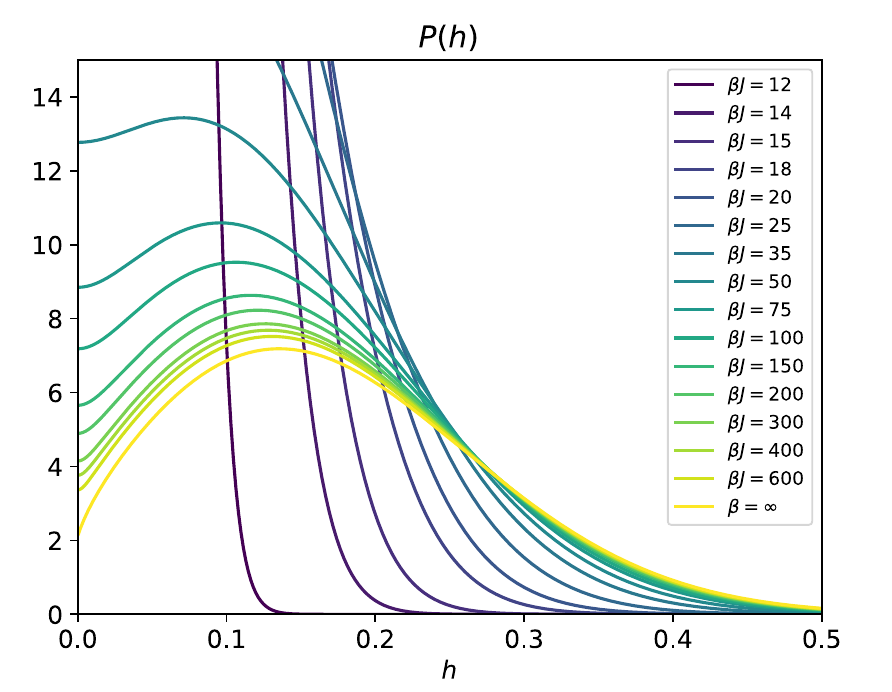}
\caption{Local field probability distribution $\mathbb{P}(h)$ for the classical Heisenberg spin glass, for a range of temperature values.}
\end{figure}

\begin{figure}
\centering
\includegraphics[scale=0.65]{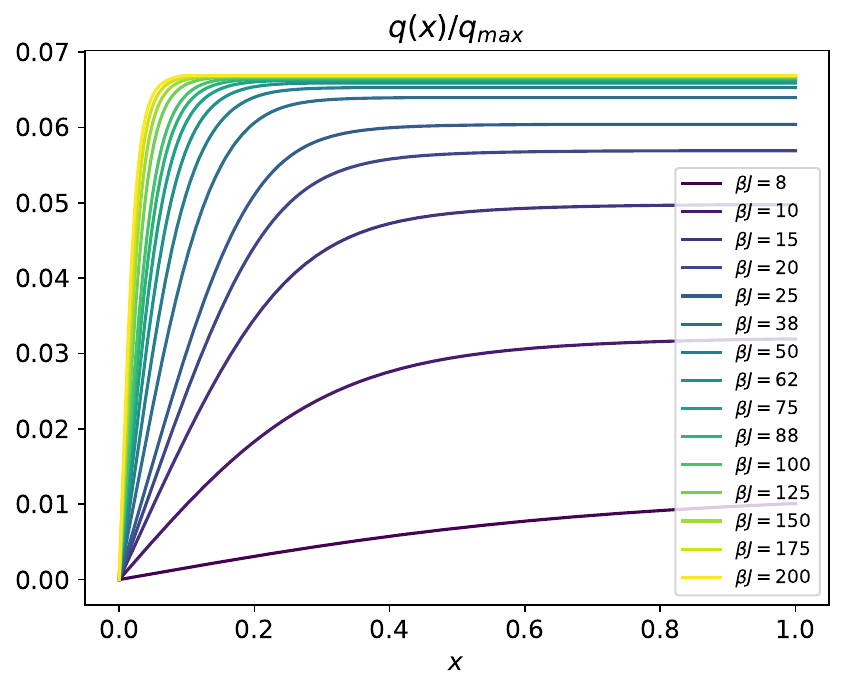}
\caption{Parisi order parameter $q(x)$ for the quantum Heisenberg spin glass, for a range of temperature values. We did not attempt to reach convergence for the breakpoint at high temperature ($\beta J < 50$), as its position has a negligible influence on the other properties of the spin glass (particularly, the scaling of $Q(\tau)$). }
\end{figure}

\begin{figure}
\centering
\includegraphics[scale=0.65]{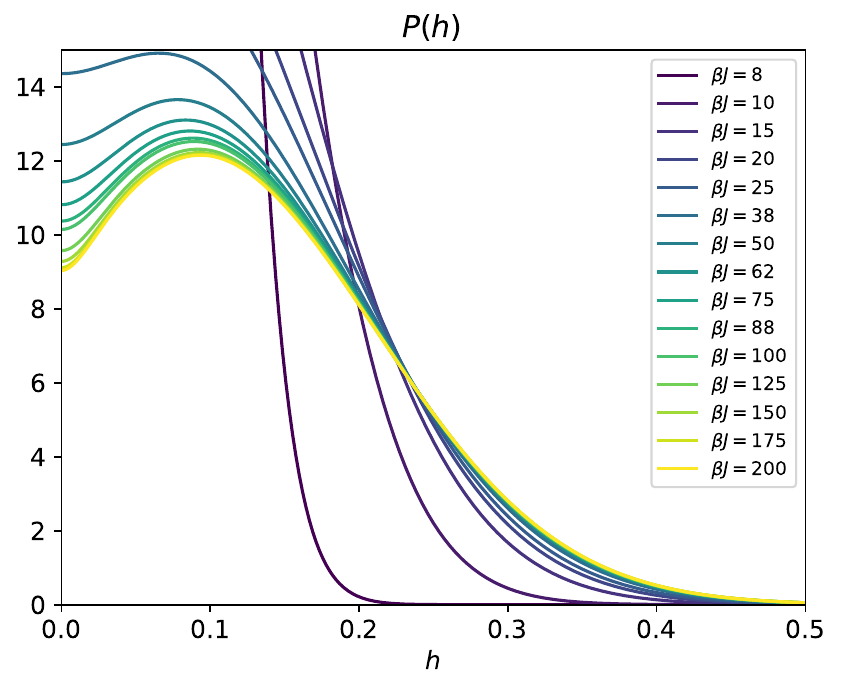}
\caption{Local field probability distribution $\mathbb{P}(h)$ for the quantum Heisenberg spin glass, for a range of temperature values.}
\end{figure}

\begin{figure}
\centering
\includegraphics[scale=0.65]{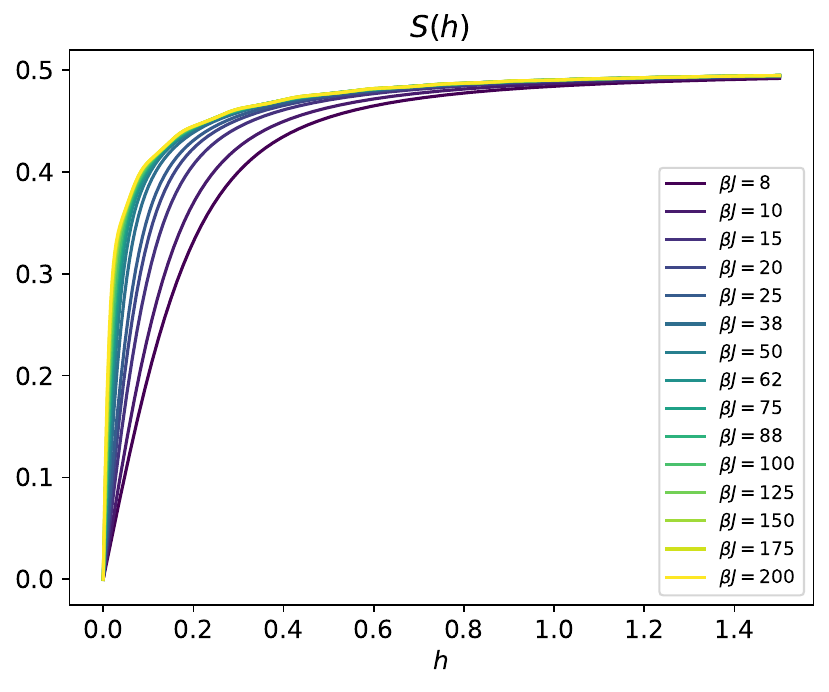}
\caption{Magnetization $\langle S \rangle_{\infty, \h}$ of the quantum Heisenberg spin glass, for a range of temperature values.}
\end{figure}

\begin{figure}
\centering
\includegraphics[scale=0.65]{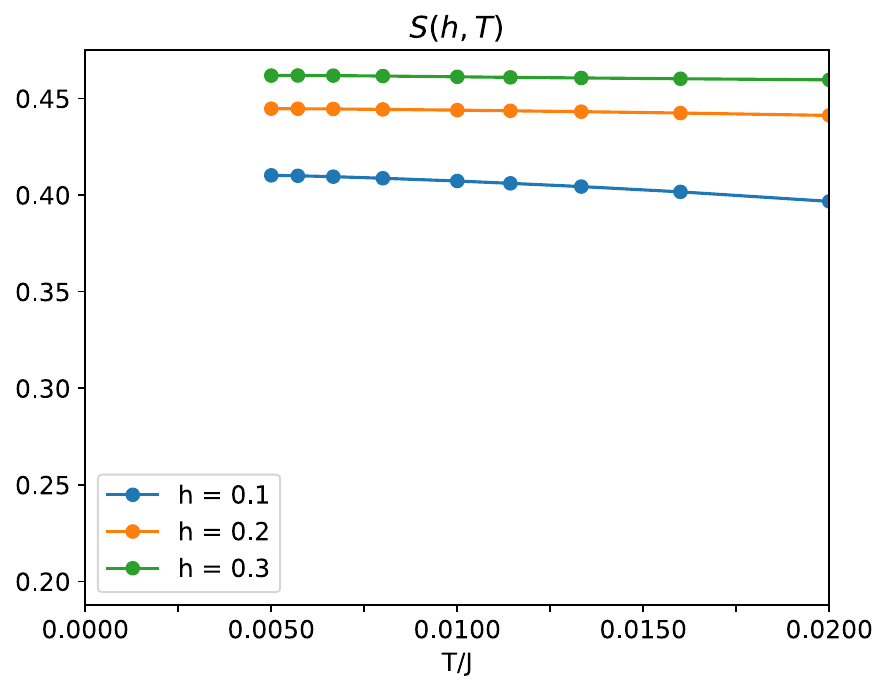}
\caption{Magnetization $\langle S \rangle_{\infty, \h}$ versus temperature, for three magnetic field values. At the lowest accessible temperatures, the magnetization is almost converged to its zero temperature shape.}
\end{figure}

\begin{figure}
\centering
\includegraphics[scale=0.65]{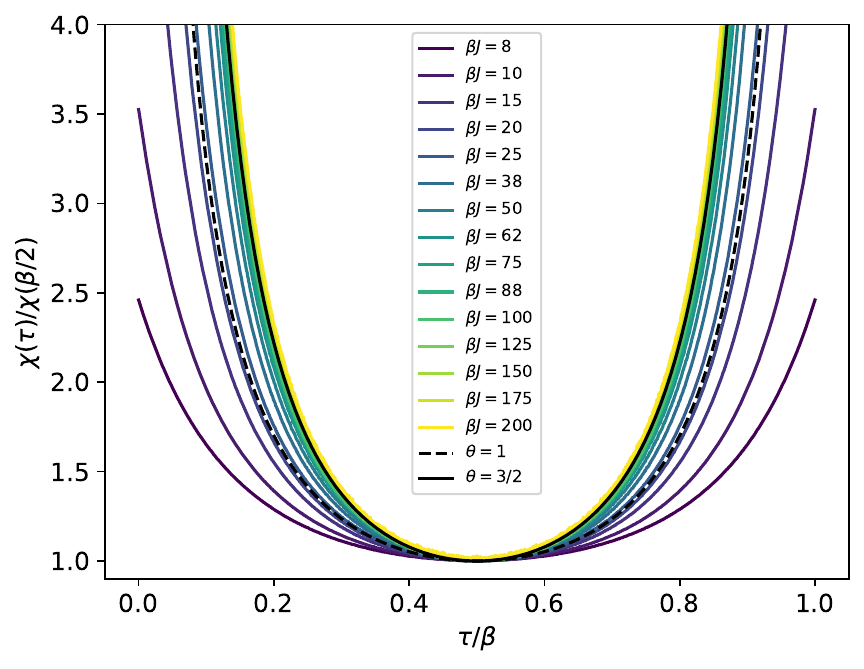}
\caption{Rescaled spin susceptibility in the quantum spin glass phase $\chi(\tau)/\chi(\beta/2)$, for a range of temperature values. It is well described by the conformal scaling form $\chi(\tau)/\chi(\beta/2) = 1/\sin(\pi \tau/\beta)^{\theta}$. The scaling functions are plotted for $\theta = 1$ and $\theta = 3/2$ (black lines).}
\end{figure}

\begin{figure}
\centering
\includegraphics[scale=0.65]{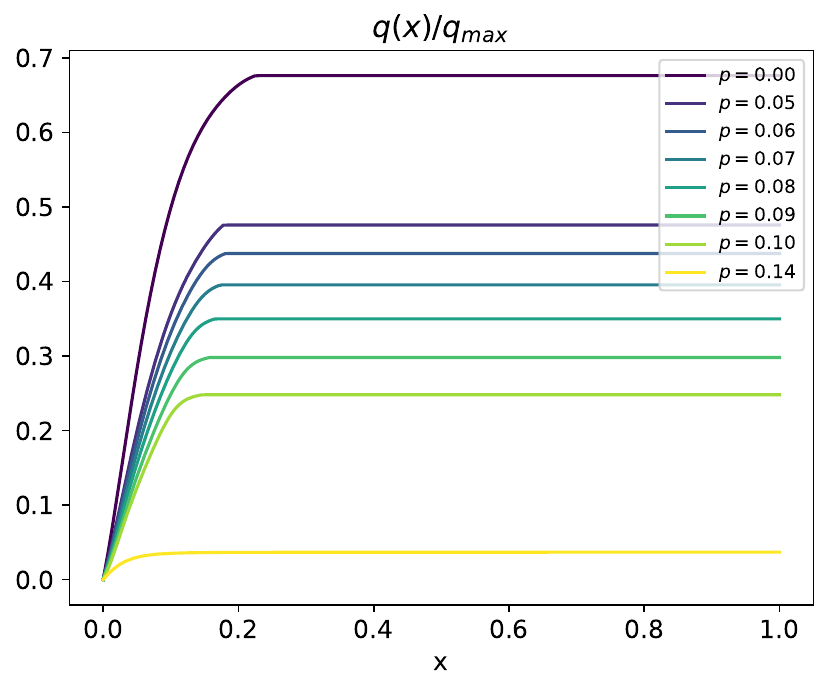}
\caption{Parisi order parameter $q(x)$ for the doped Heisenberg spin glass at $\beta J = 50$, for a range of values of the doping $p$.}
\end{figure}

\begin{figure}
\centering
\includegraphics[scale=0.65]{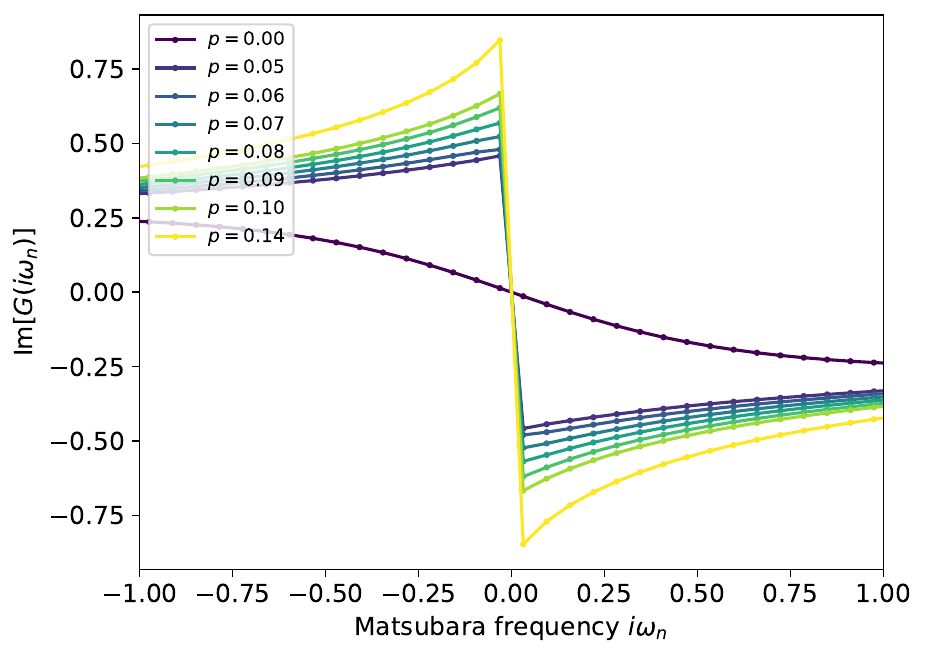}
\caption{Imaginary part of the electron Green's function for the doped Heisenberg spin glass at $\beta J = 50$, for a range of values of the doping $p$.}
\end{figure}

\begin{figure}
\centering
\includegraphics[scale=0.65]{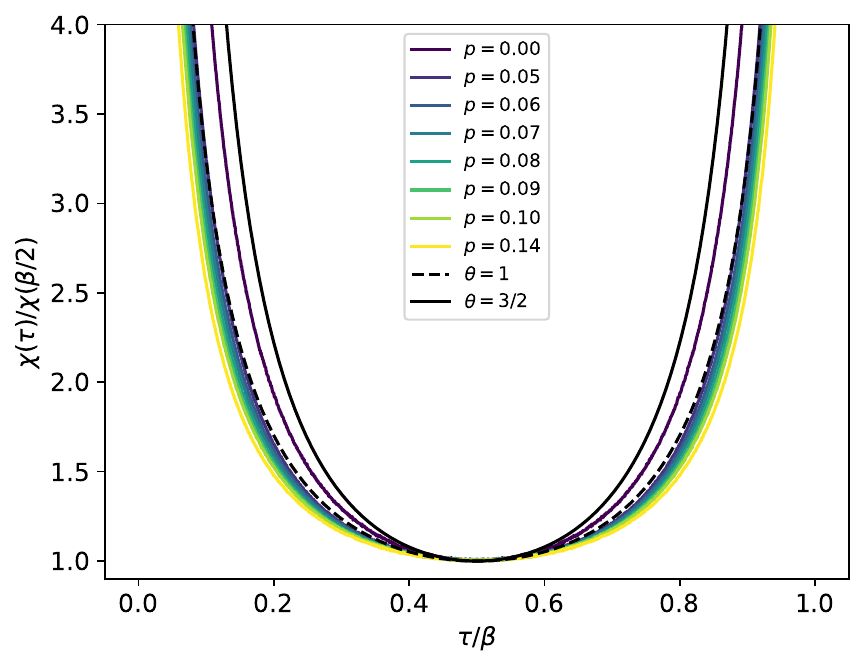}
\caption{Rescaled spin susceptibility in the metallic spin glass phase $\chi(\tau)/\chi(\beta/2)$, at $\beta J = 50$, for a range of values of the doping $p$. It is reasonably described by the conformal scaling form $\chi(\tau)/\chi(\beta/2) = 1/\sin(\pi \tau/\beta)^{\theta}$. The scaling functions are plotted for $\theta = 1$ and $\theta = 3/2$ (black lines).}
\end{figure}

\begin{figure}
\centering
\includegraphics[scale=0.65]{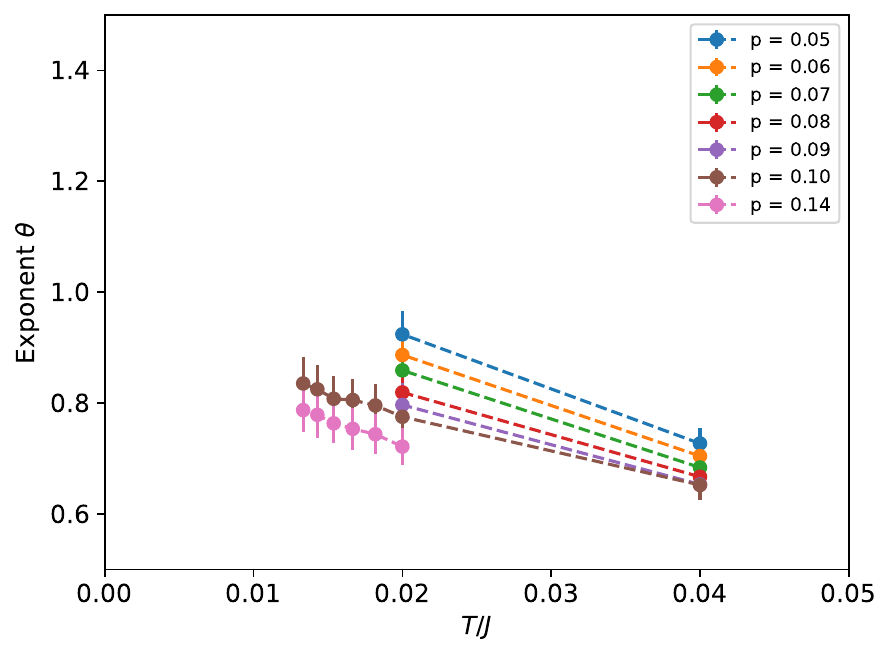}
\caption{Scaling exponent $\theta$, as a function of temperature in the metallic spin glass phase, for a range of values of the doping $p$.}
\end{figure}

\begin{figure}
\centering
\includegraphics{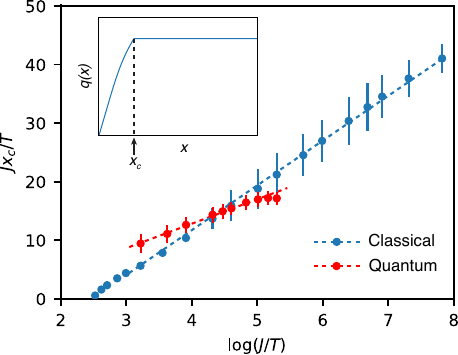}
\caption{Breakpoint $x_c(T)$ vs $T$.  $x_c(T)/T$ is a characteristic inverse energy scale governing the distribution of free energies of excited states. It scales logarithmically with $T$ in both the classical and the quantum case. Dashed lines are guides to the eye. 
Inset: definition of the breakpoint $x_c$ ($q(x>x_c)=q(1)$).}
\end{figure}

\begin{figure}
\centering
\includegraphics[width=10cm]{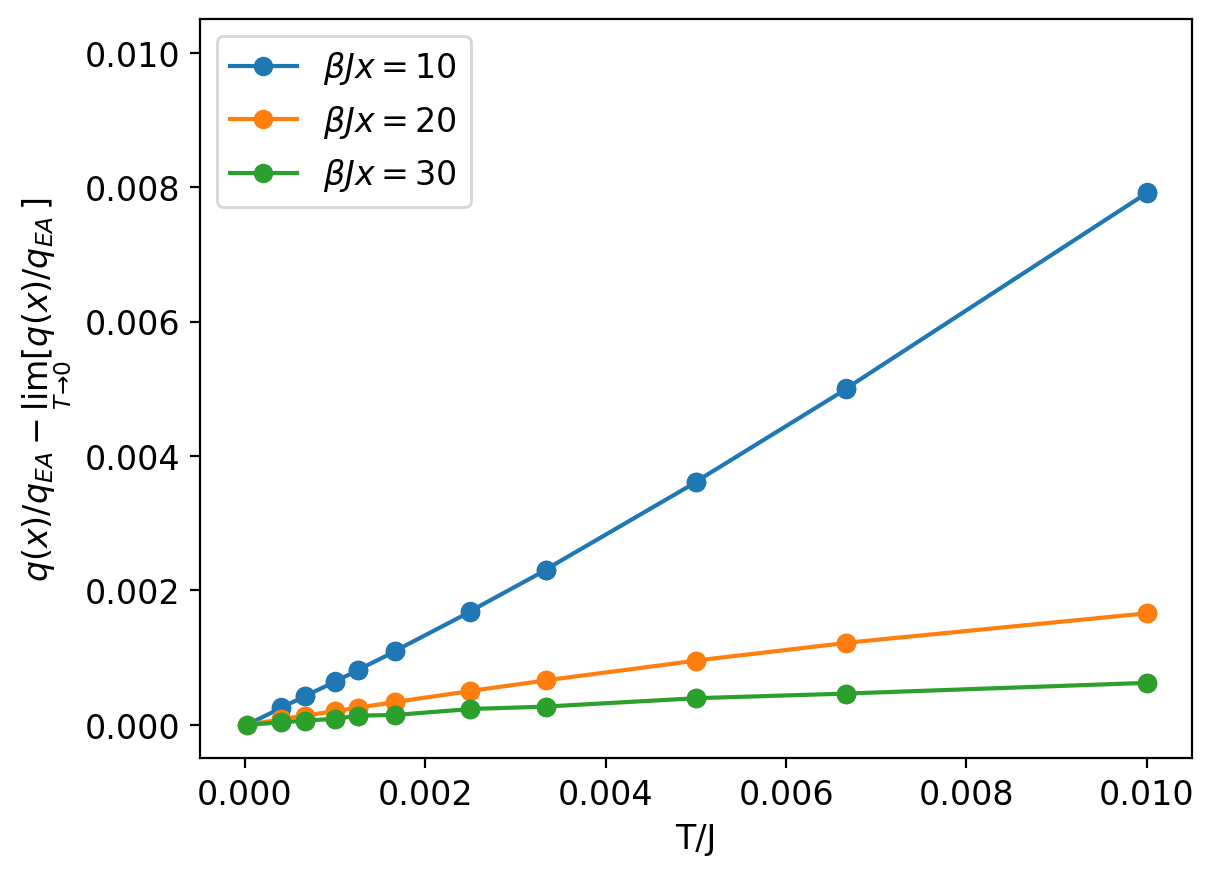}
\caption{Difference between $q(x)/q_{\rm EA}(T)$ and the scaling function $f(u = \beta x)$, versus temperature. Finite-temperature corrections to the scaling are of order $T$. }
\end{figure}